\documentclass{emulateapj}
\def\stacksymbols #1#2#3#4{\def\theguybelow{#2}
        \def\verticalposition{\lower#3pt}
        \def\spacingwithinsymbol{\baselineskip0pt\lineskip#4pt}
        \mathrel{\mathpalette\intermediary#1}}
\def\intermediary #1#2{\verticalposition\vbox{\spacingwithinsymbol
        \everycr={}\tabskip0pt
        \halign{$\mathsurround0pt#1\hfil##\hfil$\crcr#2\crcr
                \theguybelow\crcr}}}
\def\lta{\stacksymbols{<}{\sim}{2.5}{.2}}
\def\gta{\stacksymbols{>}{\sim}{3}{.5}}

\shorttitle{X-ray Isophotes in a Rotating Elliptical}
\shortauthors{Brighenti et al.}

\begin{document}

\title{X-RAY ISOPHOTES IN A RAPIDLY ROTATING ELLIPTICAL GALAXY:\\
EVIDENCE OF INFLOWING GAS}

\author{Fabrizio Brighenti\altaffilmark{1,2}, 
William G. Mathews\altaffilmark{1},
Philip J. Humphrey\altaffilmark{3},
David A. Buote\altaffilmark{3}
}

\altaffiltext{1}{University of California Observatories/Lick
Observatory,
Department of Astronomy and Astrophysics,
University of California, Santa Cruz, CA 95064 
mathews@ucolick.org}
\altaffiltext{2}{Dipartimento di Astronomia,
Universit\`a di Bologna, via Ranzani 1, Bologna 40127, Italy 
fabrizio.brighenti@unibo.it}
\altaffiltext{3}{Department of Physics and Astronomy, 
University of California at Irvine, 
4129 Frederick Reines Hall,
Irvine, CA 92697-4575}

\begin{abstract}
We describe two-dimensional gasdynamical computations of the X-ray
emitting gas in the rotating elliptical galaxy NGC 4649 that indicate
an inflow of $\sim1$ $M_{\odot}$ yr$^{-1}$ at every radius.  Such a
large instantaneous inflow cannot have persisted over a Hubble time.
The central constant-entropy temperature peak recently observed in the
innermost 150 parsecs is explained by compressive heating as gas flows
toward the central massive black hole.  Since the cooling time of this
gas is only a few million years, NGC 4649 provides the most acutely
concentrated known example of the cooling flow problem in which the
time-integrated apparent mass that has flowed into the galactic core
exceeds the total mass observed there.  This paradox can be resolved
by intermittent outflows of energy or mass driven by accretion energy
released near the black hole.  Inflowing gas is also required at
intermediate kpc radii to explain the ellipticity of X-ray isophotes
due to spin-up by mass ejected by stars that rotate with the galaxy
and to explain local density and temperature profiles.  We provide
evidence that many luminous elliptical galaxies undergo similar inflow
spin-up.  A small turbulent viscosity is required in NGC 4649 to avoid
forming large X-ray luminous disks that are not observed, but the
turbulent pressure is small and does not interfere with mass
determinations that assume hydrostatic equilibrium.
\end{abstract}

\vskip.1in

\section{Introduction}

X-ray isophotes of massive, rotating elliptical galaxies 
convey valuable information 
about the rotation of the hot, virialized interstellar gas, 
the interaction of this gas with gas recently 
ejected from evolving stars, 
and the sense of radial motion -- in or out -- of the hot gas. 
These issues are relevant to understanding the 
physical nature of the hot gas in 
galaxy groups and clusters 
in which gas loses energy by radiating X-rays 
but does not appear to cool to low temperatures as expected. 
In traditional cooling flows the radiating gas 
approximately maintains the local virial temperature 
as it flows slowly toward the 
center of the confining gravitational potential, 
then cools catastrophically near the center. 
The so-called cooling flow problem arises because there is 
insufficient 
spectroscopic evidence for cooling gas, previously cooled 
gas or recently formed stars.
This paradox is apparent in massive galaxy clusters, galaxy groups 
and individual elliptical galaxies, i.e. the cooling flow problem is 
independent of scale. 
For this reason observations of X-ray luminous, 
relatively nearby elliptical galaxies can provide 
important clues to unravel the mysteries of cooling flows 
that do not seem to cool.

The recent deep 81ks Chandra observation of
the nearby X-ray luminous elliptical galaxy NGC 4649 by
Humphrey et al. (2008)
led to the detection of
a massive central black hole ($M_{bh} = 3 \times 10^{9}$
$M_{\odot}$).
This supermassive black hole was found for the first
time directly from X-ray data by assuming that the hot gas is
very nearly in hydrostatic equilibrium.
The increased quality of the X-ray observations of NGC 4649
has motivated us to develop a dynamical model for the
hot gas flow in this rotating galaxy.

At any time the radial flow of hot gas must either be
inward, outward, a combination of the two, or stationary.
Many people 
think that cooling inflows cannot occur 
in elliptical galaxies or elsewhere because of the absence of 
X-ray spectral emission at low and intermediate temperatures 
(e.g. Xu et al. 2002).
We showed for galactic-scale cooling flows 
that dust expelled from stars can cool the gas 
so fast that very little low-temperature thermal X-ray emission occurs 
(Mathews \& Brighenti 2003a). 
Nevertheless, cooling inflows cannot dominate over time since the 
central mass accumulated would greatly exceed 
limits set by the velocity dispersion of central stars. 
Detailed gasdynamical models with {\it ad hoc} 
heating sources have shown that the transition 
from cooling inflows to strong low-density 
wind outflows is very abrupt 
(Mathews \& Brighenti 2003b). 
In the absence of extreme fine tuning,
most of the X-ray luminous hot 
gas bound to elliptical galaxies cannot be flowing 
globally outward, although inhomogeneous outflowing 
regions of lower gas density 
(and X-ray emissivity) in jets or buoyant regions are not ruled out.
If the gas is not flowing at all,  
continued enrichment by Type Ia 
supernovae for a few Gyrs will raise the iron abundance to 
several times that of the underlying stars, 
which is not generally observed 
(e.g. Humphrey \& Buote 2006).
One way to avoid this enrichment catastrophe 
is to recognize that some or most of the SNIa iron may 
cool radiatively in a way that cannot be observed, 
as discussed by Brighenti \& Mathews (2005). 
However, the modest SNIa enrichment observed 
in the hot gas can be understood 
if much of the hot galactic gas flows radially inward, 
and the abundance can be even further reduced if 
more extended circumgalactic 
group-scale gas with relatively low metallicity 
flows inward past the sources of stellar enrichment. 
Circumgalactic gas is likely since Humphrey et al. (2006)
derive a group-level virial mass for NGC 4649, 
$3.5 \times 10^{13}$ $M_{\odot}$.

Several additional observations of NGC 4649 
can be understood most easily in terms of gaseous inflow. 
First, the central gas temperature peak 
discovered by Humphrey et al. (2008) within a few 100 parsecs  
was predicted by cooling {\it inflow} models as gas
approaches a massive central black hole  
(Brighenti \& Mathews 1999). 
While only a central gas pressure peak is sufficient 
to determine the black hole mass, 
a peak in the gas temperature is a signature of 
compressional heating as hot gas flows inward 
toward a strongly concentrated mass 
or small rotating disk. 
This compressive heating can occur even when the gas 
is losing energy by standard radiative losses. 
Additional support for subsonic compressional 
heating during inflow follows from the observation 
that the entropy of hot gas 
closest to the black hole is lower than in any 
other region in NGC 4649 so it has not been recently 
heated by the central AGN.

Another important signature of subsonically inflowing gas 
seen in NGC 4649 and other similar elliptical galaxies 
is a central flattening of X-ray isophotes on kpc scales.
Elliptical X-ray isophotes on galactic scales have 
different implications than 
on larger scales which have been used to infer flattening 
of the dark halo potential (Buote et al. 2002).
Isophotal flattening in excess of that 
required by the underlying 
stellar potential can be interpreted as 
a spinning up of hot inflowing gas expected when 
angular momentum 
is exchanged with gas recently 
ejected from old, mass-losing red giant stars 
that participate in the galactic rotation.  
We review below the recent compilation 
of Chandra isophotes of elliptical galaxies
by Diehl \& Statler (2007) and show 
that they generally resemble NGC 4649 in 
central flattening and implied inflow spin-up.
NGC 4649 is an ideal candidate galaxy to study 
the influence of rotation on the X-ray isophotes because 
of its remarkably high rotation rate for a massive, 
cored E galaxy, its proximity, and 
our good fortune to have received at least one 
moderately deep Chandra observation. 

In the following discussion we review the implications of 
recent Chandra images of NGC 4649 and present gasdynamical 
models of increasing complexity. 
Unlike our previous studies of rotating cooling flows 
(Brighenti \& Mathews 1996, 2000),
we include the important contribution of inflowing 
and non-rotating 
circumgalactic gas from a group-scale halo. 
In addition, since the central 
X-ray isophotal flattening observed in NGC 4649 and other 
massive ellipticals is less than that expected 
from the conservation of angular momentum,
we invoke an {\it ad hoc} hot gas turbulence that can transport 
angular momentum away from the galactic spin axis. 
By combining turbulence and circumgalactic inflow it is 
easy to explain the isophotal flattening observed in NGC 4649. 
Moreover, the turbulent viscosity that is consistent 
with the X-ray isophotes of NGC 4649 generates a 
turbulent pressure that is much less than the gas pressure. 
Therefore, the subsonic turbulent activity required to 
reduce the angular momentum does not significantly degrade 
the total mass determinations of Humphrey et al. (2008) 
based on assuming hydrostatic equilibrium. 
We assume that the mild turbulence we require 
is produced by energy released 
near the central black hole or active galactic nucleus (AGN).

Our treatment here is done under the assumption that most 
of the {\it observed} X-ray emission in 
NGC 4649 can be explained with inflowing gas.
We do not propose that this is the complete 
solution to the gas flow problem in this or 
other similar galaxies, but only demonstrate that 
the X-ray observations are consistent with 
an {\it apparent} global inflow at the present time.
To avoid unobserved catastrophic radiative cooling 
and a huge mass concentration in the galactic core, 
we must also assume that there is a 
(possibly buoyant) return mass outflow 
that is too intermittent, 
too hot, or too cunningly disguised 
to contribute to Chandra observations 
(Mathews \& Brighenti 2004; Brighenti \& Mathews 2006).

\section{Stellar Rotation, X-ray Isophotes and Cooling Inflow}

NGC 4649 (M 60) is a luminous, E1/E2 elliptical galaxy with 
boxy, cored stellar isophotes. 
Much is known about this well-observed galaxy 
because of its high luminosity 
and proximity (15.6 Mpc: Tonry et al. 2001). 
Unlike other luminous group-centered elliptical galaxies,
NGC 4649 has an unusually large galactic rotation, 
increasing along the major axis 
to about 110 km s$^{-1}$ at 3.8 kpc. 
Because of its large line of sight rotation, we shall 
generally assume for simplicity that NGC 4649 is viewed 
nearly perpendicular to its axis of rotation. 
Like many massive elliptical galaxies, 
NGC 4649 has a pair of low power radio jets 
each extending out 3 kpc from the galactic nucleus 
(Stanger \& Warwick 1986; Shurkin et al. 2008).

The dashed line in Figure 1 
shows the major axis rotation curve $v_{rot}(r)$ of 
NGC 4649 measured by 
De Bruyne et al. (2001) and Pinkney et al. (2003).
The upper solid curve in Figure 1 shows the local circular 
velocity $v_{circ}(r)$ 
based on the combined mass of black hole, stars, and dark halo 
determined by Humphrey et al. (2008).
Also plotted in Figure 1 as rising dotted lines are trajectories
of gas that conserve the same specific angular momentum $rv$ as that
of the local stars at four observed radii.
For example if all the hot gas at 3.8 kpc ($\log r = 0.6$)
were rotating with the stars,
as this gas flows inward it would
spin up and collapse into a rotationally supported 
disk at $r_d \approx 1$ kpc where
the rising dotted line intersects $v_{circ}(r)$. 
Such a large disk, which would be extremely obvious in the X-ray 
images of NGC 4649, has not formed 
(although smaller disks are possible). 
This rotational cooling flow catastrophe could be cited 
as an argument that the hot interstellar gas 
in elliptical galaxies cannot be 
flowing inward, but such a conclusion is unwarranted 
without further examination.

\footnotetext[1]{We computed the X-ray ellipticity profile
of NGC 4649 by applying the
algorithm described in Buote et al, (2002) to the
flat-fielded 0.3--7.0 keV image taken with the Chandra ACIS-S3
chip. To prevent X-ray point-sources from distorting the ellipticity
profile, we first processed the image to remove all sources detected
with the {\it wavdetect} tool (part of the CIAO 3.4 software package),
following the procedure described in Fang, Humphrey \& Buote (2009).}

Figure 2 shows the apparent radial variation of the 
isophotal X-ray ellipticity $\epsilon_X(r)$ of 
NGC 4649 observed with Chandra by 
Diehl \& Statler (2007) 
and more deeply by Humphrey\footnotemark[1].
Some of the observational 
uncertainty within about 0.5 kpc of the galactic 
center may be due to difficulties in identifying and removing 
bright point sources 
and to the larger pixel size relative to the elliptical image 
near the galactic center.
However, the two sets of X-ray observations 
in this small central region are consistent 
within observational uncertainties 
and those of Humphrey are in good agreement with 
optical ellipticities.
Over the radial range 
$0.5 \lta r \lta 10$ kpc, $\epsilon_X$ 
increases monotonically with decreasing galactic (major axis) radius. 
We interpret this important feature of $\epsilon_X(r)$ in NGC 4649 
as a (somewhat diluted)  
spin-up flattening expected from the naive inflow argument 
in Figure 1.

It is significant that the radial variation 
of $\epsilon_X$ in Figure 2 beyond about 1 kpc 
is opposite to that of the ellipticity 
of the stellar image $\epsilon_*$ based on $R$-band observations 
of Peletier et al. (1990). 
Indeed, in the small region $0.5 \lta r \lta 1$ kpc 
($-0.3 \lta$ Log r $\lta 0$) it is likely that 
$\epsilon_X$ is larger than $\epsilon_*$, 
i.e. the hot gas is flatter than the local starlight, 
a feature seen in many massive elliptical galaxies (Diehl \& Statler 2007).
Such a trend arises naturally 
if the hot gas flows inward and is spun up to rotate 
faster than the local stars. 
In an oblate stellar ellipsoid 
having a highly concentrated de Vaucouleurs-type profile 
the gravitational potential 
is always more spherical than the local stellar density 
contours because the potential contributed by 
the dense central core acts much like a point mass.
The starlight in NGC 4649 is flattened only slightly by galactic
rotation,
i.e. $v_{circ} >> v_{rot}$ (Fig. 1).

To explore further this argument for inflowing gas
we consider a variety of alternative possibilities. 
First consider non-radiating hot gas that does not flow inward.
If NGC 4649 is assumed to have an axisymmetric oblate geometry, 
the X-ray ellipticity $\epsilon_X$ of 
non-rotating, non-radiating hot gas in hydrostatic equilibrium 
in its slightly flattened stellar distribution 
must be less than about half that of the local stars
$\epsilon_*$ if the stellar density is constant with radius 
and indeed is typically very much less 
due to its strong central density concentration 
(cf. Figure 2-13. of Binney \& Tremaine 1987).
If slowly rotating but non-radiating hot gas in such a galaxy 
has a virial temperature and specific 
angular momentum similar to that of the stars, $\epsilon_X$ will increase 
above the non-rotating case 
but certainly cannot exceed $\epsilon_*$ -- i.e. as a firm upper limit
the gas cannot rotate faster than the stars from which it was ejected.
Finally, if the hot virialized gas is
allowed to radiate and flow inward, 
its rotation velocity and $\epsilon_X$ will increase even
further until at a small enough radius 
$\epsilon_X$ may exceed $\epsilon_*$. 
This is the type of inflow spin-up that we observe in NGC 4649 
where $\epsilon_X \gta \epsilon_*$
is seen in a limited region in Figure 2.
In the presence of inflowing, low-angular momentum 
gas from the circumgalactic halo region, 
as we expect in NGC 4649, 
$\epsilon_X \gta \epsilon_*$ 
is even more difficult to achieve without inflow spin-up.
In \S 9 we review evidence for inflow spin-up in other 
nearby X-ray luminous elliptical galaxies. 

The rotational cooling inflow catastrophe 
toward large disk formation in NGC 4649 (as in Fig. 1) can be 
mitigated by the influence of extended circumgalactic 
gas and interstellar turbulence.
In view of its origin by cosmic accretion,
the virialized hot gas in 
the extended group-scale gaseous halo around NGC 4649 
probably has little or no global rotation. 
Consequently, 
the specific angular momentum of gas ejected from red giants 
rotating with the galaxy will be reduced as it mixes with 
gas of lower angular momentum flowing in from the halo.
Mixing inflowing halo gas with stellar ejecta not only 
avoids forming large X-ray disks that are not observed, 
but also maintains the gas phase iron abundance 
close to that observed. 
In addition, it is well known that turbulent viscosity 
in the hot gas can spread angular momentum 
away from the axis of rotation, 
also reducing the likelihood of large disk formation. 
Turbulence can be driven by the intermittent 
release of AGN energy associated with a small accretion 
onto the central black hole 
as evidenced for example by the radio jets observed in NGC 4649. 
When circumgalactic inflow and turbulence are combined, 
it is possible to flatten the X-ray isophotes by an amount
consistent with observation.

\section{Computational Procedure}

The gasdynamical equations that we use to 
describe rotating, non-turbulent gas flow in NGC 4649 
are identical to those 
discussed in Brighenti \& Mathews (2002) 
and Mathews \& Brighenti (2003b):
\begin{equation}
{ \partial \rho \over \partial t}
+ {\bf \nabla} \cdot \rho {\bf u}
= \alpha \rho_*
- q(T) {\rho \over t_{cool}}
\end{equation}
\begin{equation}
\rho { d {\bf u} \over d t} =
\rho \left[ { \partial {\bf u} \over \partial t}
+ ({\bf u} \cdot {\bf \nabla}){\bf u} \right]
= - {\bf \nabla} P
- \rho {\bf \nabla}\Phi - \alpha \rho_* ({\bf u} - {\bf u}_*),
\end{equation}
and
\begin{equation}
\rho {d \varepsilon \over d t}
- {P \over \rho} {d \rho \over dt} =
- { \rho^2 \Lambda \over m_p^2}
+ \alpha \rho_*
\left[ \varepsilon_o - {P \over \rho} - \varepsilon
+ {1 \over 2}|{\bf u} - {\bf u}_*|^2  \right]
\end{equation}
where $\varepsilon = 3 k T / 2 \mu m_p$ is the
specific thermal energy.
The specific rate that mass is lost from an old stellar 
population of initial mass $M_{*}$ is 
approximated with 
$\alpha_*(t) = 4.7 \times 10^{-20} (t/t_n)^{-1.3}$ s$^{-1}$
where $t_n = 13$ Gyrs (Mathews 1989). 
We assume that all of the gas ejected from evolving red giants 
eventually merges with the hot gas and C and N abundances 
in the hot gas 
provide evidence that this assumption is reasonable 
(Werner et al. 2006)\footnotemark[2].
Gas at temperature $T$ and abundance $z$ loses thermal 
energy at a rate $-(\rho/m_p)^2 \Lambda(T,z)$ erg cm$^{-3}$ s$^{-1}$. 
We adopt the cooling coefficient of 
Sutherland and Dopita (1993), 
$n_{ions}n_e\Lambda_{sd}(T,z)$ erg cm$^{-3}$ s$^{-1}$, 
where in Equation 3 
$\Lambda = [(4 - 3\mu)(2 + \mu)/25\mu^2]\Lambda_{sd}$
and $\mu = 0.61$ is the molecular weight.

\footnotetext[2]{The assumption that 
a significant fraction of the mass lost from 
orbiting, mass-losing stars 
is eventually incorporated into the hot interstellar gas 
in elliptical galaxies 
has been considered for many years (e.g. Mathews 1990; 
Mathews \& Brighenti 2003b). 
Parriott \& Bregman (2008) and Bregman \& Parriott (2009) 
provide recent, somewhat contrasting, computational 
studies of the interaction of gas ejected from mass-losing stars 
with ambient hot gas at rest.
It is likely that some fraction of the stellar ejecta 
resides briefly in small clouds of 
warm $T \sim 10^4$K gas ionized by 
the diffuse UV of old post-asymtotic branch stars. This gas 
is thought to be the source of diffuse optical line emission typically 
observed in elliptical galaxies. However, the lifetime of 
this warm gas cannot exceed about $10^6$ yrs since otherwise its 
collective optical line luminosity would exceed values typically observed 
(Mathews 1990). Because of their large surface to 
volume ratios, these small warm clouds probably merge into 
the hot interstellar gas by conductive heating.
}

As discussed above, gas cannot simply flow into the core of 
NGC 4649 and accumulate there over most of the Hubble time, 
because the total mass in and near the central black hole would be 
much larger than observed. 
Centrally inflowing gas must therefore 
flow out from the core in a manner consistent with 
the observed X-ray image, which, as we 
show here, resembles a cooling inflow. 
Since we do not treat the mass outflow in this paper,
inflowing gas in our calculations does in fact cool 
in the central grid zone or onto a disk in the $z = 0$ plane.
We consider two different ways of treating the inner boundary 
condition where gas cools: (1) allow cooled gas in 
central or disk zones to remain there at some low temperature
(we choose $T = 10^4$ K) and be supported by internal pressure or
rotation, or 
(2) remove cold gas from these zones as it begins to cool, 
allowing only hot gas to appear on the grid.
To implement this second approach we include a mass sink term 
in the continuity equation 
$q(T) \rho / t_{cool}$
where $t_{cool} = 5 m_pkT/2\mu \rho \Lambda$ 
is the local radiative cooling time at constant pressure. 
Furthermore, the dimensionless coefficient $q(T)$  
is chosen to depend very sensitively on gas temperature so that 
gas is only removed when its temperature drops below 
about $10^6$ K,
\begin{equation}
q(T) = q_{cool} \exp\left( - {T \over T_{cool}} \right)^2
\end{equation}
where $q_{cool} = 2$ and $T_{cool} = 5 \times 10^5$ K. 
Except for the central zones, 
this gas removal scheme has no effect on the flow 
where $T \gg T_{cool}$.
(Brighenti \& Mathews 2002).

The source terms
$\alpha_*\rho_* (\varepsilon_o - P /\rho - \varepsilon)$
in equation (3) represent the heating of the hot interstellar gas
of specific energy $\varepsilon$ by the mean energy
of stellar ejecta $\varepsilon_o$ less the work done
$P/\rho$ in displacing the hot gas.
The mean gas injection energy is
$\varepsilon_o = 3 k T_o /2 \mu m_p$ where
$T_o = (\alpha_* T_* + \alpha_{sn} T_{sn})/(\alpha_* + \alpha_{sn})$.
The stellar temperature $T_*$ can be found by
solving the Jeans equation,
but this term is small and it is sufficient to use
the approximation, $T_* = (\mu m_p/k){\sigma_*}^2$,
where $\sigma_*$ is the local observed stellar velocity dispersion 
in the galaxy.

Heating by Type Ia supernovae, each of energy
$E_{sn} \approx 10^{51}$ ergs, 
is assumed to be smoothly distributed in the hot gas and 
is described by multiplying
the characteristic temperature of the
mass $M_{sn}$ ejected,
$T_{sn} = 2 \mu m_p E_{sn} / 3 k M_{sn}$,
by the specific mass loss rate from supernovae,
$\alpha_{sn} = 3.17 \times 10^{-20} {\rm SNu}(t)                                         
(M_{sn}/M_{\odot}) \Upsilon_B^{-1}~~~{\rm s}^{-1}$, 
where $\Upsilon_B = 7.5$ is the assumed $B$-band 
stellar mass to light ratio.
The supernova rate is expressed in
traditional SNu-units, i.e. 
the number of supernovae in 100 yrs expected from stars of total
luminosity $10^{10}L_{B\odot}$.
In this paper we adopt a constant supernova rate 
SNu = 0.06 which is considerably lower than the 
approximate observed value for E + S0 galaxies 
SNu = 0.16 (Cappellaro et al. 1999). 
The Type Ia supernova rate of Cappellaro et al.  is so 
high that many ellipticals that are known to contain
hot virialized gas would rapidly develop winds and become 
essentially gas-free. 

For the stellar mass profile in NGC 4649 we integrate a numerical 
fit to the J-band luminosity density given in Figure 2 of 
Humphrey et al. (2008) using a mass to light ratio
$M_*/L_J = 1.43$ in solar units. 
The total stellar mass is $2.32 \times 10^{11}$ $M_{\odot}$. 
For simplicity we adopt a spherical E0 model for NGC 4649 
which should be a good approximation even in the presence 
of the observed stellar rotation which is much less 
than the circular velocity (Fig. 1). 
Also recall that the gravitational potential of 
oblate ellipsoidal elliptical galaxies 
(with centrally peaked density profiles) 
is more spherical than the local equal-density contours.
The dark NFW halo is described by a virial mass 
$M_v = 3.3 \times 10^{13}$ $M_{\odot}$ with 
concentration $c = 13.5$
(Humphrey et al. 2006, 2008). 
Finally, 
the central black hole has mass $3.35 \times 10^9$ $M_{\odot}$.
The stellar temperature distribution 
was found by assuming a virial relation 
$T_*(r) = (\mu m_p/k)\sigma_*(r)^2$ 
where the stellar velocity dispersion is a fit to the 
observations of Pinkney et al. (2003) ($r < 2.2$ kpc) and 
De Bruyne et al. (2001) ($r > 2.2$ kpc).

The major axis rotation velocity of NGC 4649 
$v_{rot}(r)$ is also 
taken from the observations of 
De Bruyne et al. (2001) and Pinkney et al. (2003) 
along the major axis 
with $v_{rot}(r)$ assumed to be constant beyond the 
most distant observation. 
Since there is very little rotation along the minor axis, 
we assume a simple cosine variation 
to describe the two dimensional 
galactic rotation, $v_*(R,z) = v_{rot}(r)cos(\theta)$ 
where $r = (R^2 + z^2)^{1/2}$, is the spherical radius, 
$\theta = \arctan(z/r)$ is the angle relative to the 
major axis plane and $v_{rot}(r)$ is the stellar rotation 
observed along the major axis.

Turbulent viscosity is essential to interpret the observed
variation of X-ray isophotal ellipticity in NGC 4649.
Since the rotational component of the gas velocity is 
typically much greater than that in the meridional plane, 
we consider only the viscous transport of the 
azimuthal gas velocity component $v_{\phi}$.
Our treatment of the viscous modification of the 
angular momentum density $\rho R v_{\phi}$ follows the 
same prescription as described in detail 
in Brighenti \& Mathews (2000) to which the reader is 
referred. 
We also include a (small) viscous dissipation term in 
the thermal energy equation as described in 
Brighenti \& Mathews (2000)\footnotemark[3].

\footnotetext[3]{ 
In this treatment we model the viscous terms directly 
as in the Navier-Stokes approximation 
(with additional energy dissipation) in which all 
terms are computed on the computational grid. 
We consider only the dominant viscosity 
contributed by the largest 
turbulent elements, approximated with a single 
parametrized viscosity. 
Recently Scannapieco \& Brueggen (2008) 
have developed an alternative computational model 
for turbulence that includes dynamical contributions 
from subgrid levels, although in this treatment 
additional unspecified coefficients must be determined by 
comparing solutions with more detailed simulations or 
laboratory experiments.
} 

We assume that the turbulent viscosity can be expressed 
as a combination of dimensionally correct factors, 
\begin{equation}
\mu(R,z) = \rho(R,z) \cdot v_t \cdot fr
\end{equation}
where $r$ is the radial coordinate, $\rho(R,z)$ 
is the computed gas density profile in 
cylindrical coordinates $R,z$, 
$v_t$ is a characteristic turbulent velocity and 
$fr$ is the characteristic mean free path of the largest 
turbulent eddies 
where $f < 1$. 
Since these parameters appear only as a product,
only one parameter $v_tf$ characterizes each solution.  
The objective is to find the minimum viscosity 
$\mu(r)$ consistent with the observed X-ray ellipticity 
(Fig. 2).

In principle the problem we address is somewhat ill-posed 
because of (1) unavoidable multi-Gyr transients 
and (2) the intrinsic time variability of important parameters.
Regarding the first point, suppose (as we usually do) 
that the initial hot gas galactic atmosphere 
has the observed radial gas density and temperature profiles 
but is at rest with no rotation.
As rotation is induced by stellar mass loss
(from the term $\alpha \rho_* ({\bf u} - {\bf u}_*)$ in eq. 2)
the interstellar gas spins up and 
an inward flow develops because of radiative losses.
These transient adjustments 
can take many Gyrs.
Consequently, we have decided to (rather arbitrarily) select 
6 Gyrs as a time to view all models. 
As we describe below, a few flows have nearly reached 
a steady state by this time while others may still be 
slowly evolving. 
Secondly, the stellar mass loss rate and supernova rate 
are expected to change with time as the galactic stellar population 
evolves.
However, to include this change introduces additional uncertainties 
and time dependences that we believe complicate our assessment of 
the gas flow velocity in present-day elliptical galaxies.
For this reason we have decided to keep the stellar mass loss 
rate and 
Type Ia supernova rate fixed at their adopted current 
values for all flow solutions: 
$\alpha_* = 4.70 \times 10^{-20}$ s$^{-1}$ and SNu = 0.06. 

We also monitor the approximate iron abundance in the gas 
as the calculation proceeds using an additional 
continuity equation for iron,
\begin{equation}
{\partial (\rho z) \over \partial t}
+ {\bf \nabla}\cdot (\rho z){\bf u}
= \alpha_* \rho_* z_*
+ \alpha_{sn} \rho_* {y_z \over M_{sn}} \xi
\end{equation}
where $\xi = \rho / \rho_H$ and $z$ is the iron abundance 
relative to hydrogen, i.e. $z = \rho_{Fe}/\rho_H$.
The local gas phase 
iron abundance $z$ appears in the radiative cooling coefficient
$\Lambda(T,z)$, but this has very little influence on the 
overall gas dynamics. 

As mass is ejected from evolving stars, its iron abundance 
is assumed to contribute instantaneously to the local gas. 
For this purpose we adopt a simple stellar iron profile 
\begin{equation}
z_*(r) = z_{*0} \left[ 1 + \left({r \over r_{*c}}\right)^2\right]^{-0.15}
\end{equation}
typical for large elliptical galaxies 
(e.g. Arimoto et al. 1997).
In this last equation the stellar iron abundance is 
expressed in solar units, i.e. 
$z_{\odot} = 1.82 \times 10^{-3}$.
When circumgalactic gas is initially present, we
assume that it has a uniform iron abundance
0.6$z_{Fe,\odot}$ (see \S 7).

\section{One-dimensional Cooling Flows\\
and Numerical Strategy for Gas Removal}

We begin with a discussion of simple one-dimensional, 
non-rotating solutions to the 
equations above in the presence of circumgalactic gas.
The initial density profile for the circumgalactic gas 
is determined by integrating the equation of 
hydrostatic equilibrium toward larger radii, 
normalizing the central density to 
obtain good agreement with the density and temperature 
profiles observed by 
Humphrey et al. (2006; 2008) within 20 kpc and 
assuming isothermal gas ($T = 10^7$ K) beyond this radius.
These one-dimensional solutions are useful in selecting a preferred  
mass sink term $q(T)$ for numerically removing cooled gas. 

We create computational grids of different resolutions 
by selecting the central grid zone 
size $\Delta r$ and successively increasing the zone size by 
a factor $(1 + \varepsilon)$. 
Figure 3 shows the radial gas density and temperature profiles
after computing for 6 Gyrs on a spherical 1D grid, 
using three grid resolutions $\Delta r = 15$, 150 and 600 pc
with $\varepsilon = 0.0568$, 0.03196 and 0.01508 respectively, 
from left to right in Figure 3.
Each solution is presented with two different 
numerical procedures 
to treat cooling gas: $q = 0$ for which the centrally 
cooled gas remains in the computational grid (light solid lines)
or $q = q(T)$ as in equation (4) where cooled gas is 
continuously removed after each computational time step
(heavy solid lines).

The left pair of panels in Figure 3 
shows number density and temperature 
profiles computed with the highest spatial resolution.
For these flow solutions   
the gas temperature has a small positive gradient 
throughout most of the central region.
Positive $dT/dr$    
often occurs in classical cooling flows when circumgalactic gas 
is present because the virial temperature in the dark halo 
exceeds that of the stellar potential alone 
(Mathews \& Brighenti 2003b).
However, within about 100 pc of the galactic center 
the temperature increases due to compression 
as the gas approaches the central black hole, 
a feature which is absent if the black hole is removed 
(cf. heavy dotted profiles in Fig. 3). 
This central temperature rise -- for inflowing gas -- 
is an essential attribute 
of massive black holes (Brighenti \& Mathews 1999) 
and was used by Humphrey et al. (2008) to verify the presence 
of the central black hole in NGC 4649.
However, as gas flows even closer to the black hole, 
it ultimately cools by radiative losses 
in a very small region inside the central temperature peak. 
Clearly, some physical process 
(that we do not include here) must intervene 
in this small region to avoid an excessive 
concentration of of cooled gas. 
In any case, the central temperature rise in $r \lta 100$ pc 
is responsible for the high gas temperatures observed by 
Humphrey et al. in their central aperture. 

For $r \gta 30$ pc 
the high resolution solutions 
with $\Delta r = 15$ pc are in good agreement 
with each other regardless of 
the numerical treatment of cooling gas,
$q = 0$ or $q = q(T) \ne 0$.
Note that the high resolution 
temperature profile in Figure 3 beyond about 1 kpc is 
slightly lower than 
the observed temperature at all radii. 
(The temperature profile computed in the 
two-dimensional version of this flow 
agrees better with the observations
presumably because the flow velocity is 
less constrained in two dimensions.)
These high-resolution one-dimensional flows are unable to 
reproduce the small observed negative gas temperature gradient 
observed in the region $0.2 \lta r \lta 1.5$ kpc 
($-0.7 \lta$ Log$r \lta 0.2$)
which is largely unaffected by the potential of the central black hole. 
(This feature occurs naturally in two-dimensional flows.) 
The observed gas density is fit nicely beyond about 1 kpc 
in these flows, but continues to rise 
above the observed density inside this radius. 
This is a standard shortcoming of all simple one-dimensional 
cooling flows as discussed in detail by 
Mathews \& Brighenti (2003b).

The central and right panels in Figure 3 
show the results of identical one-dimensional 
calculations performed with coarser grid resolutions, 
$\Delta r = 150$ and 600 pc.
The central temperature peak produced by the 
black hole potential is not resolved in 
either of these calculations, but the computed temperature 
profiles now depend on the choice of $q$.
When $q = 0$ (light lines) the accumulated cold gas resides in a 
sphere of radius $\Delta r$ which produces an 
unphysical negative temperature gradient in adjacent central 
grid zones. 
This may be due to a small compression as the 
subsonic inflow flows toward 
the barrier presented by a sphere of dense cooled 
gas in the central zone.
When $\Delta r = 150$ pc 
this feature is seen only near $r \approx 300$ pc, 
but when $\Delta r = 600$ pc
the temperature gradient is negative out to about 3 kpc. 
However, when cooled gas is continuously removed 
with $q = q(T)$ (heavy lines), the computed 
(positive) temperature gradient is in essential agreement 
at all resolutions $\Delta r$. 
As a result of this numerical experiment,  
we shall remove cooling gas 
using the expression for $q(T)$
in equation (4) in all two-dimensional calculations 
where 
$\Delta R$ and $\Delta z$ are typically much larger than 15 pc.


\section{Two-Dimensional Flows: Individual Influence of 
Circumgalactic Gas and Turbulence}

\subsection{Without Circumgalactic Gas or Turbulence}

To verify that a rotational cooling flow catastrophe 
could in fact occur in NGC 4649, 
we begin our two dimensional calculations with a 
non-turbulent flow in which 
galactic stars are the only source of 
hot gas, i.e. we do not include group-scale gas at any radius. 
The computed flow after 6 Gyrs is shown in the left panels of 
Figure 4.
Since stellar mass loss and supernova rates are held constant, 
as discussed above, this solution reaches a steady state 
after only $\sim$2 Gyrs 
and remains essentially unchanged afterwards.
Gas ejected from stars is heated to the galactic 
virial temperature 
by gravitational compression and by 
the source terms in equation 3. 
The top left panel in Figure 4 shows that the computed 
(and azimuthally averaged) gas 
density (solid line) is much less than that observed 
by Humphrey et al. (filled circles) 
and this discrepancy increases with galactic radius beyond 10 kpc.
This density shortfall arises because the stellar mass loss rate 
is unable to replenish gas at the same rate that it cools 
by radiative losses and flows into the cooling disk 
near the galactic core. 
The computed gas temperature is also much cooler than the 
observations because the local virial temperature 
associated with the stars $\sim T_*$ is less than that in 
the surrounding dark halo. 
The lower left panel in Figure 4 
shows an image of the computed flow viewed in 
bolometric X-ray emission.
The intense X-ray emission that results 
as gas cools toward the rotationally supported disk of radius 
$\sim1.8$ kpc 
dominates the entire X-ray appearance of the galaxy. 
These computed X-ray isophotes have little resemblance to those 
observed in NGC 4649 (dashed lines)
or observations of any other known elliptical galaxy. 
This verifies our prediction of a large X-ray disk 
for NGC 4649 in the absence of circumgalactic gas 
and turbulent viscosity.

\subsection{With Turbulence but Without Circumgalactic Gas}

The central panels in Figure 4 show flow profiles
and isophotes computed for NGC 4649 after 6 Gyrs
when gas viscosity is included but again 
without circumgalactic gas so stellar mass loss is 
the only source of gas.
Turbulence in this flow is described with parameters 
$v_t = 50$ km s$^{-1}$ and $f = 0.1$ as defined in equation (5). 
The spatial variation of $\mu(R,z)$ 
can be seen in Figure 5 where we plot $\mu(R,0)$ and $\mu(0,z)$ 
for $v_t = 50$ km s$^{-1}$ and $f = 0.05$ 
(parameters appropriate for the ``reference model'' discussed below). 
The density profile in this solution is improved 
but is still too low overall because of the inability of 
old galactic stars to resupply gas 
as it flows inward by radiative losses.
The temperature profile is marginally improved but is still 
much lower than observed by Humphrey et al. (2008). 
However, the X-ray isophotes for this model are greatly 
improved and are in good agreement with observation, 
considering the uncertainties in the observed 
ellipticity $\epsilon_X(r)$ shown in Figure 2. 
Because of observational uncertainties 
in $\epsilon_X(r)$ within about 0.5 kpc, 
it is not possible to compare computed 
and observed isophotes in this central region. 
But it is clear that subsonic turbulence must be an essential 
component in successful flow models for NGC 4649.

\subsection{With Circumgalactic Gas but Without Turbulence}

The gas density in NGC 4649 is not expected to drop to zero 
just beyond 20 kpc, the most distant density observation 
by Humphrey et al. (2008).
In all subsequent computed gas flows discussed here we 
adopt an initially stationary gas density profile 
that matches the observations of NGC 4649 
out to 20 kpc but is extrapolated further by solving 
the equation of hydrostatic equilibrium for an isothermal 
gas ($T = 10^7$ K) at rest in the potential of 
the stellar and dark halo mass described above. 
The initial density profile extrapolated 
in this manner beyond 
20 kpc has the same logarithmic slope as 
the observations between 1 and 20 kpc.
While the gas density is low in this extrapolated region,
the total mass of this gas exceeds that 
observed within 20 kpc.

The gas flow computed 
with circumgalactic gas but without viscosity is shown in 
the panels on the right in Figure 4.
The gas density and temperature profiles 
are now in much better agreement with observations 
than previous flows without circumgalactic gas.
After 6 Gyrs the inflowing circumgalactic gas 
continues to create density and temperature profiles that 
agree with observations of NGC 4649 within 20 kpc. 
Indeed, this agreement is another indirect argument 
for cooling inflow. 
(The temperature downturn 
in the central grid zone is an artifact 
of incomplete removal of cooling gas at this 
moment in the calculation.)
Beyond about 1.5 kpc, 
the computed X-ray isophotes are in reasonably good 
agreement with observations.
However, in the range $0.5 \lta r \lta 1.5$ kpc  
the computed X-ray isophotes are still 
much flatter than those observed in NGC 4649. 
We conclude that 
circumgalactic gas improves the agreement 
with observations but cannot by itself 
provide completely satisfactory flow solutions. 

The reduced prominence of the X-ray disk when circumgalactic 
gas is included 
occurs because we assume that the initial circumgalactic gas 
is not rotating. 
Non-rotating gas flowing toward NGC 4649 from 
the group-scale halo 
provides an inertial brake on 
gas ejected from evolving stars that 
systematically share the galactic rotation. 
But spin-up and disk formation 
still occur at a level that is inconsistent 
with observation (lower right panel in Fig. 4). 
If we had allowed for the time-dependent decrease in 
the stellar mass loss rate expected 
as a single stellar population evolves, 
${\dot M}_* \propto t^{-1.3}$, 
the size of the cooling disk would be even larger 
than shown at the bottom right in Figure 4 
because of the increased mass of high-angular momentum 
gas introduced in the past.
In this sense our central disk sizes computed with 
constant $\alpha_*$ are conservative. 

Because of inadequacies in the solutions shown in 
Figure 4, we now explore whether 
the formation of large, X-ray detectable disks 
can be avoided by considering both 
the turbulent redistribution of angular momentum 
and rotational suppression 
by mixing with non-rotating circumgalactic gas. 

\section{2D Flows with Both Turbulence and Circumgalactic Gas}

We now describe three representative evolutionary gas flows 
in which the initial density and temperature profiles 
are extrapolated into the circumgalactic halo 
beyond the observations as described above. 
All gas is assumed to be initially at rest, 
but it acquires a small radial inflow due to 
radiative losses and angular momentum 
is supplied by mass ejected from evolving stars 
in the rotating galaxy. 

\subsection{Reference Model}

The left panels in Figure 6 show gas density 
and temperature profiles and X-ray isophotes 
for NGC 4649 computed after 6 Gyrs with viscosity parameters
$v_t = 50$ km s$^{-1}$ and $f = 0.05$ 
for this ``reference model''.
At this time the computed temperature and density are in good agreement 
with observed profiles. 
The central temperature peak due to the black hole 
is not resolved at the grid resolution used, 
(central $\Delta R = \Delta z = 150$ pc).
Furthermore, at this resolution 
the flow near the very center is artificially disturbed 
when gas cools rapidly in the central zones. 
In the left panels we show typical flow excursions at three 
times. 
The X-ray isophotes beyond $\sim0.5$ kpc 
also follow the observed $\epsilon_X(r)$ 
within acceptable uncertainties. 
Because of possible uncertainties in $\epsilon_X(r)$
within 0.5 kpc -- both observational (Fig. 2) and 
computational -- 
it is not possible to make a detailed comparison of our 
flow with observations in this important region.
However, the central rotation has not completely disappeared in 
this flow and gas continues to cool into a small disk 
of radius $\sim$300 pc. 
It is significant that 
this disk is similar in size to the small dusty disks discovered 
with the Hubble Space Telescope in the cores of many 
elliptical galaxies (e.g. Lauer et al. 2005). 
With deeper Chandra observations the 
gas flowing into these small disks may become visible in X-ray images.

Rotation is helpful in achieving a better agreement 
with azimuthally averaged gas density and temperature profiles 
in the left panels of Figure 6. 
The rotational flattening visible in the computed  
X-ray isophotes near the core of this 
flow (lower left panel) causes the azimuthally averaged gas density to 
flatten similar to the observations of NGC 4649.
In addition, rotation introduces a shallow negative 
temperature gradient in the range $0.2 \lta r \lta 1$ kpc 
($-0.7 \lta$ Log$r \lta 0$)
that is unrelated to the black hole potential. 
This arises because of compressional heating as gas flows 
toward the disk.
Neither the central density flattening or the shallow 
negative $dT/dr$ within $\sim 1$ kpc appear in otherwise similar 1D 
spherical flows.

The turbulent velocity $v_t = 50$ km s$^{-1}$ in this flow 
is much less than the sound speed in the hot gas 
in NGC 4649 beyond about 1 kpc,
$c_s \approx 460$ km s$^{-1}$.
As a result, the ratio of turbulent to gas pressure 
$P_t/P = (v_t/c_s)^2 \approx 0.012$ is very small,  
provided the viscosity scale factor 
$f$ exceeds $(v_tf)/c_s \approx 0.005$, which we think is likely. 
For a Kolmogorov spectrum 
the velocity of turbulent elements $v$ varies with their size 
$\ell$ as $v \propto \ell^{1/3}$ 
so $\mu(\ell) \propto v\ell \propto \ell^{4/3}$ 
and most of the momentum is carried by the larger eddies; 
this argues against adopting very small values of $f \sim \ell/r$ 
and $\ell$
for a fixed turbulent velocity at any radius $r$.
Determinations of the underlying 
mass by Humphrey et al. (2008) 
using hydrostatic equilibrium implicitly require 
$P_t/P \ll 1$ which is consistent with the level of turbulence 
required to remove unobserved large X-ray disks in NGC 4649.
Finally we note that the agreement with observed 
X-ray isophotal ellipticity could be improved with a more 
highly parameterized and finely-tuned function for $\mu(R,z)$, 
but we do not pursue this alternative here.
Altogether, we regard this as a successful model and 
refer to it subsequently as the ``reference model''.

While the agreement between our 2D reference model 
and X-ray observations of 
NGC 4649 is quite satisfactory, it is useful to keep in mind 
some of the approximations we have made: 
NGC 4649 is assumed to be viewed perpendicular to 
its rotation, we use a cosine law for $v(R,z)$ to interpolate 
between observations on the major and minor axes, 
we assume that the rates of stellar mass loss and supernova 
explosions are constant during the calculation, 
we compare with observations only after 6 Gyrs, 
our removal of cooled gas introduces intermittency 
in the central zones,  
our turbulent viscosity $\mu(R,z)$ is parameterized 
in a simple way, and viscosity is applied only to the 
azimuthal shear velocity. 
Consequently, exact agreements with the observations are 
not expected.

\subsection{Additional Models}

One of our objectives is to estimate the minimum 
turbulent viscosity and pressure necessary to bring the X-ray isophotes
into agreement with X-ray observations of NGC 4649. 
Models with $v_tf$ less than that of the reference model 
tend to resemble the inadequate flow for NGC 4649 
without turbulence (but including circumgalactic gas) 
shown in Figure 4.
For example, the central panels of Figure 6 show the flow 
after 6 Gyrs 
computed with $v_t = 25$ km s$^{-1}$ and $f = 0.025$.
The gas density and temperature profiles in this flow 
are similar to the reference model, including 
excursions due to sudden cooling.
However, the computed X-ray isophotes are unacceptably flat 
near $r \sim 1$ kpc.
Our overall subjective assessment of this model is that it 
is less successful than the reference model which 
can be regarded as an acceptable model having the 
smallest turbulent viscosity.

Finally, in the right panels of Figure 6 we explore the 
effect of a turbulent viscosity that exceeds that of the 
reference model, i.e. $v_t = 50$ km s$^{-1}$ and $f = 0.1$.
The X-ray isophotes are seen to be considerably rounder 
for this flow and are a reasonable match to the observations 
beyond about 0.5 kpc from the center.
However, for this solution the flow approaches that of a 
non-rotating pure cooling flow with density that is too high 
(for $r \lta 0.7$ kpc) and with a temperature gradient 
that remains positive in the range 
$0.2 \lta r \lta 1$ kpc. 
Consequently, we regard 
this model as less successful overall than the reference model
for NGC 4649.

\section{Approximate Gas Abundance Profiles}

The gas phase iron abundance profile in NGC 4649 
depends on the Type Ia supernova rate (SNu = 0.06), 
the stellar abundance profile $z_*(r)$ 
(given by equation 7 with parameters
$z_{*0} = 1.5z_{\odot}$ with solar abundance 
$z_{\odot} = 1.82 \times 10^{-3}$, 
$r_{*c} = 100$ pc), 
the assumed (constant) iron abundance in 
the circumgalactic gas $z_{cgg} = 0.6z_{\odot}$, 
and the time elapsed since the beginning of the calculation.
Figure 7 shows hot gas
iron abundance profiles computed to 6 Gyrs for two
flows discussed above. 
The heavy solid line shows 
$z_{Fe}/z_{\odot}$ with inflowing 
circumgalactic gas and with reference 
model viscosity parameters 
($v_t = 50$ km s$^{-1}$ and $f = 0.05$).
In general the abundance profile tends to flatten
slightly with increasing turbulent viscosity, 
but we do not illustrate these details.
This abundance profile agrees rather closely with
the abundance data in Figure 7 based on observations 
described in Humphrey et al. (2007).

The thin line in Figure 7 shows the abundance profiles 
for the same flow parameters but without circumgalactic gas.
In flows without circumgalactic gas 
after 6 Gyrs the gas abundance approaches 
$\sim 3.6$ times solar with or without turbulent viscosity.
Since gas-phase iron abundances this high 
are rarely if ever observed, 
Figure 7 provides dramatic evidence 
that low-abundance circumgalactic gas must 
flow slowly in from the extended halo, 
as in a normal cooling flow, diluting the enrichment 
received from galactic supernovae.

\section{Higher Computational Resolution}

In Figure 8 we show a high resolution 2D 6-Gyr calculation of the 
reference flow in which the grid is uniform
for $R,z < 5$ kpc, with $\Delta R = \Delta z = 33$ pc 
(i.e.  $\sim150$ uniform zones in $R$ and $z$).  
Then the grid extends for another 150 zones in both directions 
with increasing zone size to reach the outer boundary at 190 kpc.
At this higher resolution 
the gas density and temperature profiles 
follow the observations within 1 kpc even more closely than 
with normal resolution (left panels in Figure 6). 
However, the central temperature peak is not as sharp as in the 
1D high resolution flow in Figure 3. 
This is probably due to unrealistically large 
(subsonic) meridional velocities in 
the $R,z$-plane near the center -- our current 
viscosity prescription only 
damps the azimuthal component of the gas velocity. 
The bolometric X-ray image is essentially indistinguishable from 
the reference flow in Figure 6. 
It is important to note that 
our flows at high resolution do not oscillate 
unrealistically near the center as mass is removed 
by cooling. 
This is probably due to the smaller mass that is removed 
from smaller central zones 
that causes relatively less disturbance in nearby zones.

The top panels in Figure 8 show the influence of 
viscosity on the density and temperature profiles.
As the viscosity increases, the temperature profile becomes 
flatter and the density profile steepens.
This sensitivity of the central temperature profile in 
$-0.5 \lta$ Logr $\gta 0$ ($0.3 \lta r \gta 1$ kpc) 
is a measure of the role of viscosity in rotating, 
quasi-disk like flows in this region. 
This is an interesting detail that should be explored 
in other X-ray luminous elliptical galaxies. 
As before, identical flows without a central black hole 
show no central temperature peak inside 300 pc.

The bottom panels in Figure 8 shows the 
X-ray isophotes and the 
gas velocity field in the high-resolution reference flow 
superimposed on the local X-ray isophotes.
The snake-like flow in the central few kpc is characteristic 
of all our solutions 
and also occurs in flows that are not rotating.
We believe that this non-radial flow pattern may result 
as the inflow adjusts to accommodate 
gas recently ejected from stars that enters the flow 
with a radial gradient that is steeper than the 
local gas density profile in hydrostatic equilibrium. 
These velocities have spatial scales that are  
similar to those observed in optical emission lines 
in luminous elliptical galaxies 
(Caon, Machetto \& Pastoriza 2000), 
but the observed velocities are somewhat higher. 
The velocity field varies with time, 
causing the X-ray profiles to 
distort and wander slightly from those 
at 6 Gyrs shown in Figure 8.
The X-ray ellipticity can be estimated with 
$\epsilon_X = 1 - z_i/R_i$ where $z_i$ and $R_i$ 
are the intercepts of X-ray isophotes on the $z$ and $R$ axes 
in Figure 8. 
These approximate ellipticties, shown as crosses in Figure 2, 
are in good agreement with observed values.
While X-ray observations in Figure 2 indicate that 
$\epsilon_X > \epsilon_*$ near 0.6 kpc ($\log r = -0.2$),
the observations of Humphrey (open circles) suggest that 
$\epsilon_X \approx \epsilon_*$ closer to the origin. 
This near equality could arise if viscous damping 
increases more strongly toward the center
than in our computed flows, allowing a nearly
spherical flow of gas onto the central black hole in NGC 4649. 
The crosses in Figure 1 show the tangential gas velocity along the
major ($z$) axis for the high resolution flow in Figure 8.
Because of inflow spin-up, the gas rotates faster than the stars for 
$r \lta 2$ kpc.

\section{Chandra Evidence of Rotational Inflow in Other Elliptical Galaxies}

In their recent analysis of Chandra X-ray images observed 
in 54 elliptical galaxies 
Diehl \& Statler (2007) found no convincing correlations between the 
X-ray appearance and the rotation of the galactic stars or 
the shape of the stellar potential. 
Specifically, they found no clear relationship between 
the mean X-ray ellipticity and line of sight stellar rotational
velocity, both averaged over 0.6 - 0.9$R_J$ where $R_J$ is the J-band
effective radius of the galactic stars.
Furthermore, 
their plots of the X-ray ellipticity 
$\langle \epsilon_X(0.6-0.9R_J) \rangle$ against 
either the stellar ellipticity 
$\langle \epsilon_*(0.6-0.9R_J) \rangle$ or the stellar rotation 
$\langle v_{rot}(0.6-0.9R_J) \rangle$ show large scatter 
without an obvious correlation.
(They do not consider the effective potential of the X-ray gas 
in its locally rotating frame.) 
These results led Diehl \& Statler to conclude that 
the hot gas in elliptical galaxies is quite far from 
hydrostatic equilibrium and that rotation cannot be the dominant 
factor that produces the hot gas X-ray ellipticities they observe. 

However, we suspect that conclusions based on 
X-ray ellipticities and stellar properties, 
when spatially averaged over the broad radial range
0.6 - 0.9$R_J$, may be insensitive to regular trends in the 
X-ray and stellar ellipticity profiles apparent 
in the unaveraged, raw data 
(Fig. 5 of Diehl \& Statler 2007).
For example, for NGC 4649 Diehl \& Statler find that the X-ray ellipticity
$\langle \epsilon_X(0.6-0.9R_J) \rangle = 0.08 \pm 0.03$
is less than the corresponding stellar ellipticity
$\langle \epsilon_*(0.6-0.9R_J) \rangle = 0.18 \pm 0.01$, 
and, for similar galaxies in their complete sample, 
Diehl \& Statler infer that the hot gas is not rotating or is
insensitive to the flattened stellar potential
in this annular bin.
However, we interpret the negative slope 
$d\epsilon_X/dr$ in NGC 4649 (Fig. 2)  
beyond about 0.5 kpc ($0.14R_J$) as clear evidence 
of rotational
spin-up resulting from a global inflow of the X-ray emitting gas.
Rotational spin-up of the hot interstellar gas 
cannot be accurately inferred from absolute values 
of ellipticity or the line of sight velocity at any particular radius. 
Instead, in the mitigating presence of 
turbulence and/or circumgalactic gas, 
rotational spin-up can only be detected by the radial trend of 
X-ray isophotal flattening, $d\epsilon_X/dr < 0$,  
particularly when $\epsilon_X > \epsilon_*$ 
near the center of the galaxy, implying that the  
gas is rotating faster than the local stars.

To explore our interpretation further, we examine the detailed 
Chandra data of 
$\epsilon_X(r)$ and $\epsilon_*(r)$ for 36 
well-observed galaxies in Figure 5 of 
Diehl \& Statler (2007).
We do not consider six of these galaxies 
(NGC 193, 383, 507, 1316, 1553, and 4526)
since their morphological classification parameter $T$ 
from LEDA exceeds -4.0, suggesting that 
they are E/S0 or later,  
although our conclusions are 
independent of this sample refinement.
With this reduced sample of 30 E galaxies, 
we subjectively classified the innermost slope 
$d\epsilon_X/dr$ into positive (+), indeterminate (0) and negative (-).
We find the number of galaxies in each slope category to be 
(+,0,-) = (4,10,16), showing that a majority of Chandra-observed 
E galaxies have negative $d\epsilon_X/dr$ as expected from inflow
spin-up\footnotemark[4].

\footnotetext[4]{Galaxies with -: IC 1262, 1459, NGC 720, 
1399, 1404, 1407, 4125, 4261, 4374, 4406, 4552, 4636, 4649,
4697, 5846, 6482; with 0: IC 4296, NGC 741, 1132, 1549, 1600, 3923,
4472, 5018, 7052, 7618; with +: NGC 315, 533, 1700, 5044.}

However, some of these galaxies have been observed at only 1 - 3 
radii within $R_J$ either because of their large distance 
and reduced spatial scale or because of the 
relatively low total number of Chandra photons available.
To allow for this, we consider a smaller subsample of 16 E galaxies, 
including NGC 4649, 
in which $\epsilon_X$ has been measured at four or more radii $\le R_J$.
For this well-observed subsample we find overwhelming evidence for 
rotational spin-up in $r \lta R_J$: (+,0,-) = (2,2,12). 
Those few 
galaxies with indeterminate or positive $d\epsilon_X/dr$ may 
have spin axes that are closer to our line of sight or 
their ordered rotation may have been masked by 
transient kinematical activity following a recent AGN outburst --
such asymmetric AGN disturbances in elliptical galaxies
have been studied and quantified in detail by Diehl \& Statler
(2008a).

In addition, for the 12 E galaxies showing clear evidence of 
inflow spin-up, $d\epsilon_X/dr < 0$, the stellar and X-ray 
(major axis) position angles are in good agreement 
either near the galaxy center or closer to $R_J$ where most of the 
stellar mass loss and angular momentum may be deposited.
Finally, of the 12 galaxies with strong evidence of 
inflowing rotational spin-up 
($d\epsilon_X/dr < 0$), 
the innermost measured X-ray ellipticity in 
10 galaxies exceeds the ellipticity of local stars. 
Since the stellar gravitational potential 
in giant elliptical galaxies 
is always more spherical than 
the local ellipsoidal stellar density distribution, 
this ellipticity excess 
cannot be understood as non-rotating gas
nearly in hydrostatic equilibrium 
in the flattened stellar potential 
or gas rotating with the local stars.
Instead, $\epsilon_X > \epsilon_*$ 
indicates that the inner hot gas in these 12 galaxies is 
rotating faster than the local stars, 
as a natural consequence of inflowing spin-up.

Nevertheless, the absence of any clear correlations among 
broadly binned averages, 
$\langle \epsilon_X(0.6-0.9R_J) \rangle$ and
$\langle \epsilon_*(0.6-0.9R_J) \rangle$ or
$\langle \epsilon_X(0.6-0.9R_J) \rangle$ and
$\langle v_{rot}(0.6-0.9R_J) \rangle$,
as found by Diehl \& Statler (2007) in their complete sample,
is interesting and should be better understood.
We believe that several factors combine to explain 
the scatter in these plots.
The mass of inflowing circumgalactic, group-scale gas 
increases with the mass of the surrounding dark halo 
which is highly variable 
among optically similar elliptical galaxies 
(Mathews et al. 2006b).
As a result, the fractional contribution of 
inflowing (non-rotating) circumgalactic gas in the hot 
gas near the galactic effective radius is expected 
to vary considerably. 
For a given rotation of the stellar system, 
the X-ray ellipticity should decrease inversely with 
the local fraction of circumgalactic gas. 
This introduces a stochastic variation 
in $\langle \epsilon_X(0.6-0.9R_J) \rangle$ 
that is unrelated to inflow spin-up. 
Secondly, we have shown here that the X-ray ellipticity 
and its radial variation depends somewhat on large 
scale subsonic velocity fields in the hot gas.
These velocities are weakly time-dependent and may 
depend on the radial distribution (and age) of mass-losing 
stars which may vary among the galaxies in the sample considered 
by Diehl \& Statler.  
Finally, the unknown turbulent viscosity profile 
in the hot gas may alter $\epsilon_X(r)$ in complicated 
and unpredictable ways because of occasional energy outbursts 
associated with AGN feedback from the central black hole.
During the period of transient
recovery following each AGN outburst, 
the systematic spatial and kinematic patterns in 
the hot gas are likely to be significantly disturbed. 

\section{Summary of Observational Evidence for Cooling Inflow}

Throughout the preceding discussion we mentioned several 
arguments why 
the X-ray emitting gas in NGC 4649 -- and in other 
massive elliptical galaxies -- appears to be 
undergoing inflow. 
In view of the remarkable implications of these observational 
and theoretical arguments, 
it is useful to summarize them again here.

\vskip0.1in
\noindent
(1) {\it Increasing ellipticity of X-ray isophotes toward the galactic 
center, often exceeding that of local stars}:
We have shown that this can arise naturally from angular momentum spin-up 
of inflowing gas driven by mass ejected from stars 
that collectively follow the global galactic rotation. 
In many large elliptical galaxies the central X-ray ellipticity 
exceeds that of the stars, indicating that the hot gas 
cannot be rotating with the stars in their potential, 
but must have a faster rotation due to inflow spin-up.

\vskip0.1in
\noindent
(2) {\it Central peak in gas temperature due to compression as gas 
flows in toward a centrally concentrated potential}: 
Temperature peaks occur as inflowing gas approaches a massive 
black hole (Fig. 3) or flows onto 
a small central disk orbiting the black hole 
(e.g. the reference model in the left panels of Figure 6). 
This temperature increase is a consequence of adiabatic heating 
as gas approaching the central black hole is compressed 
by the steepening potential. 

\vskip0.1in
\noindent
(3) {\it Dusty, rotationally supported nuclear disks}: 
Dusty disks are observed in the cores of 40 - 50\% 
of luminous elliptical galaxies (e.g. van Dokkum \& Franx 1995;
Lauer et al. 2005).
Disks of this type are expected to form naturally from dust 
ejected by 
mass-losing red giant stars within 
$\sim$1 kpc from the galactic center (Mathews \& Brighenti 2003a).
The absence of disks in other similar ellipticals suggests that 
AGN energy has disrupted them on time scales of $10^7 - 10^8$ yrs,
consistent with the spatially extended dust observed by 
Temi, Brighenti \& Mathews (2007).
Indeed, highly transient, chaotically arranged dust clouds 
are observed in in the cores of 
many ellipticals that do not contain disks.

\vskip0.1in
\noindent
(4) {\it Gas metal abundance}: 
Even though the hot gas inside elliptical galaxies 
is continuously enriched by 
Type Ia supernovae and mass-losing stars, 
the metal abundance observed in the hot gas 
is rarely much greater than that of the local stars.
This can be understood by two aspects of inflow: 
(i) dilution of the hot gas metallicity 
by low metallicity gas flowing in 
from circumgalactic, group-scale halos 
and (ii) continuous removal of metal-rich gas on galactic scales.
This latter removal could be simply a continuation of the 
circumgalactic cooling inflow but ultimately this metal-rich 
gas must cool or be transported far out into the hot gas in jets 
or buoyant regions. 
In principle, the iron from Type Ia supernovae
may cool by radiative losses before it thermally merges with the
hot gas; the main difficulty with this 
is the uncertain thermal conductivity
imposed by the unknown local magnetic field topology 
(Brighenti \& Mathews 2005).

\vskip0.1in
\noindent
(5) {\it Gas density and inflowing circumgalactic gas}: 
For galaxies like NGC 4649 the expected 
stellar mass loss rate from an old stellar population is insufficient
to maintain the relatively high gas densities observed
in the presence of inflow (e.g. Fig. 4). 
Without circumgalactic gas, 
the computed gas density becomes increasingly lower 
than the observed density at larger galactic radius.
In addition, 
gas lost from red giant stars virializes to temperatures that 
are lower than those observed in NGC 4649,  
as seen beyond about 2 kpc in the left and central panels 
of Figure 4. 
However, both of these difficulties can be 
corrected by including a cooling inflow from
circumgalactic gas virialized 
in the dark matter halos surrounding elliptical galaxies.

\vskip0.1in
\noindent
(6) {\it Similarity of observations and computed cooling flows}: 
The hot gas density and temperature profiles observed in massive
ellipticals as well as the hot gas isophotes 
can be explained with gasdynamical models of inflowing, 
radiatively cooling gas (Figs. 6 \& 8). 
The failure to observe cooling gas at intermediate temperatures 
can be understood by rapid dust-assisted cooling (Mathews \& Brighenti
2003a), provided the cooing rate does not greatly exceed 
$\sim1 M_{\odot}$ yr$^{-1}$. 
However, the time-integrated mass of cooled gas 
expected in elliptical galaxy cores is not observed, 
so the mass inflows we compute here cannot be sustained 
throughout the evolution of NGC 4649 or other 
massive ellipticals. 
Our cooling inflow model is 
a valid description of the {\it instantaneous} 
flow currently observed in NGC 4649. 
To be valid over cosmic times, 
it is necessary to  
consider the release of AGN accretion energy 
in the galactic core and associated outflows of energy and mass. 


\section{Mass Accretion Rate -- Fate of Inflowing Gas}

Figure 9 shows the rate ${\dot M}_{cool}(t)$ 
that gas cools in several flow models discussed here. 
In the absence of circumgalactic gas,
the galactic cooling rate becomes nearly equal to the 
total galactic stellar mass loss rate, ${\dot M}_{cool} \approx 0.3$
$M_{\odot}$ yr$^{-1}$ 
following an initial transient adjustment.
The computed cooling rate
is essentially independent of the turbulent viscosity.
When circumgalactic gas is included, ${\dot M}_{cool}(t)$ 
is somewhat larger $\sim0.9$ $M_{\odot}$ yr$^{-1}$. 
This cooling rate is inconsistent with the much lower value  
${\dot M}_{cool} \approx 0.05$ $M_{\odot}$ yr$^{-1}$ 
allowed by X-ray spectra and FUSE observations of 
the OVI 1032-1038\AA  
~doublet in NGC 4649 (Bregman \& Athey 2003). 

This failure to observe cooling and cooled gas is the 
cooling flow problem. 
While the X-ray properties of NGC 4649 resemble a typical 
(weakly-rotating) 
cooling flow down to $r \sim150$ pc, the innermost radius 
observed by Humphrey et al., 
there is no spectral evidence for cooling 
and no young stars are detected.
The ultra-short (gas) radiative cooling time (6 Myr) and 
the (non-rotating) 
gas flow time to the center (0.5 Myr) at radius 
$r = 150$ pc concentrate and enhance the  
AGN feedback-cooling flow problem in NGC 4649 
unlike that of any other known galaxy, group or cluster.

The currently preferred paradigm for solving the cooling 
flow problem, as explained in detail in the recent review by 
McNamara \& Nulsen (2007), is that the hot gas is being heated 
by the creation of X-ray cavities or by the dissipation 
of outwardly propagating shock waves. 
According to this hypothesis, 
as the hot gas is heated in this manner, 
it experiences little or no systematic flow 
in either radial direction.
But neither of these heating mechanisms seems likely in NGC 4649.  
Moreover, 
X-ray cavities formed with cosmic rays result not in heating 
but in global cooling due to an expansion of the 
entire gaseous atmosphere (Mathews \& Brighenti 2008).
Of course, cavities formed with ultra-hot (but non-relativistic) 
gas can increase the 
global thermal energy of the hot gas if the cavity 
gas mixes with the ambient, pre-cavity gas. 
In any case, no cavities larger than about 125 pc 
in radius are currently 
observed in NGC 4649 (Humphrey et al. 2008).
Heating by wave dissipation occurs mostly in the central 
regions where the hot gas density gradient is flatter.  
Because of this, 
as wave heating continues for several Gyrs. 
the central gas becomes much hotter than that observed 
(Mathews, Faltenbacher \& Brighenti 2006a). 

Nevertheless, if the hot gas in NGC 4649 is being heated somehow 
by AGN energy so there is no cooling inflow, 
this heating must be nearly continuous 
in time and exquisitely fine-tuned with galactic radius 
to maintain the appearance of a normal cooling inflow
complete with a central, constant-entropy temperature peak near the 
massive nuclear black hole in which the gas has a cooling time of 
only a few million years. 
However, the effect of inflow spin-up on the X-ray isophotes 
observed in NGC 4649 and other similar galaxies 
is clearly inconsistent with non-cooling flow models 
that stop the inflow with heating at every radius.
Finally, we note again that the hot gas observed within 
the central elliptical galaxy must move radially 
to avoid extremely large iron abundances, 
assuming that Type Ia iron thermally merges 
into the hot gas -- this is another difficult constraint 
when the cooling flow problem is solved with simple heating scenarios 
in which there is no radial flow of gas. 

Most of the gas that cools in our computed flows 
(with circumgalactic gas) cannot have simply contributed to 
the mass of the central black hole in NGC 4649 since its mass 
would be reached after only $\sim3$ Gyrs. 
It is also unclear how much cooling is occurring since rapid 
dust-assisted cooling (Mathews \& Brighenti 2003a) 
would make cooling in elliptical galaxies 
difficult to detect in X-ray spectra. 
Certainly, the low apparent cooling rate in NGC 4649 estimated from 
UV line emission, $\sim0.05$ $M_{\odot}$ yr$^{-1}$, 
cannot be reconciled with the larger mass inflow rates in the 
models we compute here. 

Since the X-ray evidence for global inflow is so compelling 
in NGC 4649, the most likely fate of the inflowing gas 
is that gas is being transported outward in a manner that 
is intermittent or not easily visible in X-ray images and spectra 
(Mathews, Brighenti \& Buote 2004). 
Outward mass transport may occur either 
in jets (Brighenti \& Mathews 2006) or in low density, 
buoyant outflows created for example with cosmic rays 
(Mathews \& Brighenti 2008). 
Like many massive elliptical galaxies, NGC 4649 has 
a pair of low luminosity FRI radio jets.
These jets may transport thermal gas from deep within the 
core of NGC 4649, 
provided they are similar to the mass-carrying radio jets 
found in Seyfert galaxies by Whittle \& Wilson (2004).
Some of the cooled gas and dust may also be 
intermittently removed from deep 
within NGC 4649 by jets or buoyant outflows as 
implied by far-IR dust emission observed out to 5-10 kpc 
in many elliptical galaxies
(Temi, Brighenti \& Mathews 2007). 
It is likely that the mass outflow in NGC 4649 is episodic
and that we are viewing NGC 4649 at a quiescent moment 
between radio lobe outflow events when 
gas is undergoing a cooling inflow toward the center 
(e.g. Diehl \& Statler 2008b).

\section{Conclusions}

We have shown that the X-ray emitting gas associated with the 
bright elliptical galaxy NGC 4649 has density and temperature 
profiles that can be understood as a rotating, 
radiatively cooling inflow with a mild turbulent viscosity.
Evidence of inflow is present at every radius in NGC 4649. 
On the largest scales, 
a radiatively cooling inflow of circumgalactic gas is required to dilute 
the hot gas iron abundance acquired from supernovae 
down to observed values, 
assuming that the iron ejected by Type Ia supernovae does not cool
by radiative losses before it merges with the ambient hot gas.
The relatively high gas density and temperature observed in 
NGC 4649 beyond $\sim$10 kpc can also be explained with inflowing
circumgalactic gas.
On intermediate galactic scales, 
observations of the X-ray ellipticity in 
NGC 4649 and other bright elliptical 
galaxies show that the hot gas is spun up by mass ejected from 
evolving stars that rotate collectively with the galaxy. 
Inflow spin-up is seen in the increasing X-ray isophotal  
ellipticity with decreasing galactic radius 
until it exceeds the ellipticity of the local stars 
near the galactic center.
However, for NGC 4649, which rotates unusually fast for 
a massive boxy elliptical, a small turbulent viscosity is required 
to avoid forming multi-kpc X-ray disks that are 
not observed. 
The shallow negative temperature 
gradient inside $\sim1$ kpc in NGC 4649 is X-ray evidence 
of an inflow that compresses toward a sub-kpc disk. 
The observed X-ray isophotes can be matched with a turbulent 
viscosity for which the corresponding turbulent pressure 
is much less than the gas pressure, 
so the integrated mass profiles found by assuming 
hydrostatic equilibrium (Humphrey et al. 2008) are unaffected.
This alleviates the concern expressed by Diehl \& Statler (2007) 
that the hot gas in elliptical galaxies is very far 
from hydrostatic equilibrium.
Galactic mass determinations based on X-ray thermal emission 
may fail for some elliptical galaxies  
where non-thermal pressure dominates 
the galactic core, such as NGC 4636 
(Brighenti \& Mathews 1997; Baldi et al. 2009),
but this problem does not seem to occur in NGC 4649.
On the smallest scales observed with Chandra,
within about 100 parsecs, 
the central temperature peak in NGC 4649 
discovered by Humphrey et al. (2008) is a natural consequence of 
subsonic cooling inflow. 
In spite of ongoing radiative losses, 
the central gas temperature increases 
due to (nearly adiabatic) compression as gas approaches 
the central black hole or the small disk around it. 
The kpc-scale negative temperature gradient in NGC 4649 
formed as gas compresses toward a small disk is 
not caused by recent AGN heating since the gas entropy in 
this region increases with galactic radius.
This type of central inflow is also consistent with 
nuclear disks of cooled dusty gas having radii of a few
hundred parsecs commonly observed in the cores of 
luminous elliptical galaxies. 

NGC 4649 represents the most concentrated 
known example of the cooling flow problem. 
The observed density and temperature profiles can be explained 
at every radius 
with a radiatively cooling inflow. 
At the smallest observable radius in NGC 4649, about 150 pc, 
the times for cooling and flow to the core are 
only a few million years. 
For this reason it will be very difficult to devise 
heating scenarios that maintain the gas at rest, 
perfectly mimicking a cooling inflow. 
No ideally heated model in which the hot gas in NGC 4649 remains 
stationary can account for commonly observed X-ray 
ellipticity profiles $\epsilon_X(r)$ that 
require inflow spin-up.
 
The current mass inflow rate in our successful calculations is  
$\sim1$ $M_{\odot}$ yr$^{-1}$, but the stellar mass loss 
rate was certainly several times larger in the past.  
Consequently, if this flow continued 
over a Hubble time, the total mass of cooled gas 
accumulated in the galactic core of NGC 4649 would exceed the mass 
of the black hole 
and stars within $\sim1$ kpc by factors of 3 - 10.
Therefore, the mass inflow indicated by the central 
temperature peak must be removed from within 
$\sim150$ pc 
at a (time-averaged) rate comparable to $\sim1$ $M_{\odot}$ yr$^{-1}$  
and transported out to a large radius without 
interfering with the X-ray appearance of NGC 4649 
which resembles a traditional cooling flow. 
This mass outflow may occur 
in the bipolar jets observed in NGC 4649 and many other 
bright ellipticals.
Alternatively, nuclear hot gas may become buoyant by intermittent 
local heating or by cosmic rays and flow out subsonically 
upstream in the cooling flow 
(e.g. Mathews \& Brighenti 2008). 
If the buoyant gas density is only slightly less than that 
of the ambient gas, it may not significantly disturb the radial 
gas density and temperature profiles set by the inflowing gas. 
Alternatively, 
if the density of buoyant (or jet) gas were considerably lower, 
its X-ray emission would be less easily observed against the 
brighter emission from denser inflowing gas.



\vskip.1in
\acknowledgements
Studies of the evolution of hot gas in elliptical galaxies
at UC Santa Cruz are supported by NSF and NASA grants 
for which we are very grateful.

\clearpage

\begin{figure}
\vskip3.in
\centering
\includegraphics[bb=250 216 422 769,scale=.8,angle=270]{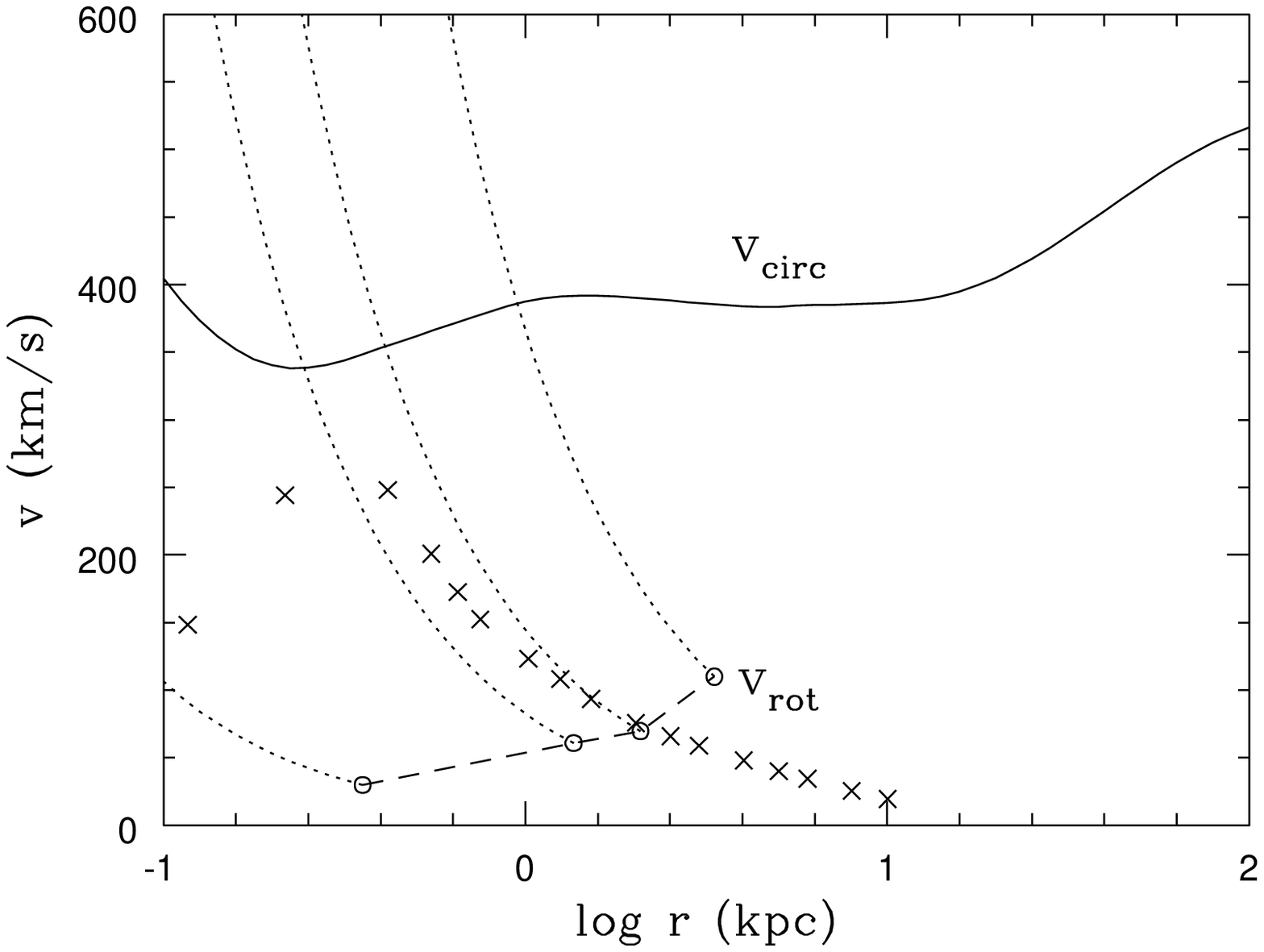}
\vskip.7in
\caption{
Comparison of circular velocity $v_{circ}$ in NGC 4649
({\it solid line})
with major axis stellar rotation $v_{rot}$
({\it dashed line})
from Pinkney et al. (2003). Dotted lines show trajectories of
constant specific angular momentum.
The crosses show the rotational velocity of the gas 
along the major axis as computed 
in our high resolution flow shown in Figure 8.
}
\label{f1}
\end{figure}

\clearpage
\begin{figure}[ht]
\vskip-2.0in
\centering
\includegraphics[bb=250 216 422 769,scale=1.0,angle=0]{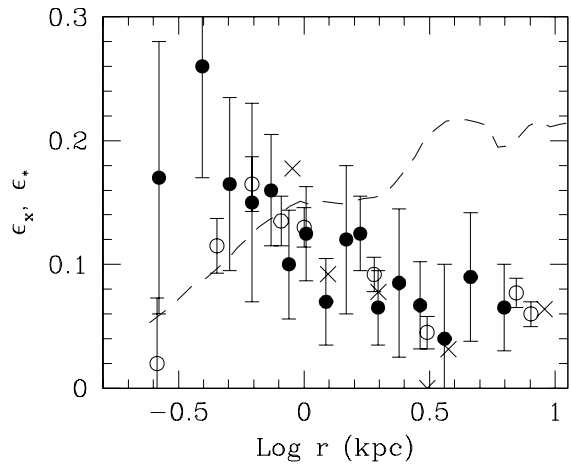}
\vskip.7in
\caption{
Ellipticity $\epsilon_X(r)$ of X-ray isophotes along major axis 
of NGC 4649 observed by
Diehl \& Statler (2007) ({\it filled circles})
and Humphrey et al. ({\it open circles}). 
Because of their deeper observations, the data from Humphrey et al.  
have smaller error bars.
The $R$-band stellar ellipticity profile 
$\epsilon_*(r)$ is shown with a dashed 
line (Peletier et al. 1990).
The crosses are approximate ellipticties of our high resolution 
rotating flow shown in Figure 8.
}
\label{f2}
\end{figure}

\clearpage
\begin{figure}[ht]
\vskip-2.0in                                                
\centering
\includegraphics[bb=200 216 372 769,scale=1.0,angle=0]{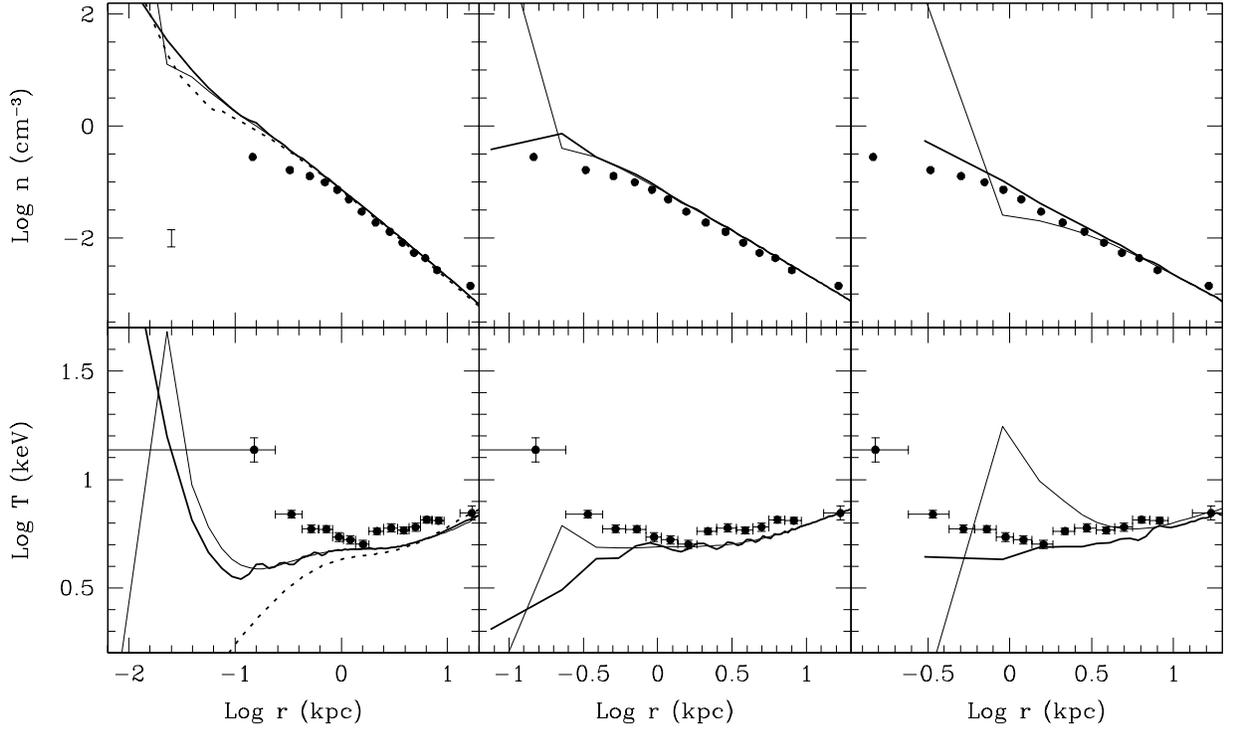}
\caption{
Electron density and temperature profiles in one-dimensional 
flows for NGC 4649 after 6 Gyrs compared with observations 
of Humphrey et al. (2008). 
A representative mean error bar for the density observations 
is shown in the upper left panel.
Calculations are repeated at three different grid resolutions 
with central zones of size $\Delta r = 15$, 150 and 600 pc 
respectively from left to right. 
Each calculation is performed with no removal of cooled gas 
($q = 0$; {\it thin lines}) and continuous removal 
($q = q(T)$; {\it thick lines}).
For comparison 
the heavy dotted lines in the left panels show  
high resolution density and 
temperature profiles with $q = q(T)$ 
when the central black hole is removed. 
The density inversion within 30 pc 
in the central panel is a transient 
feature associated with this mass removal. 
}
\label{f3}
\end{figure}

\clearpage
\begin{figure}[ht]
\vskip-2.0in
\centering
\includegraphics[bb=250 216 422 769,scale=1.0,angle=0]{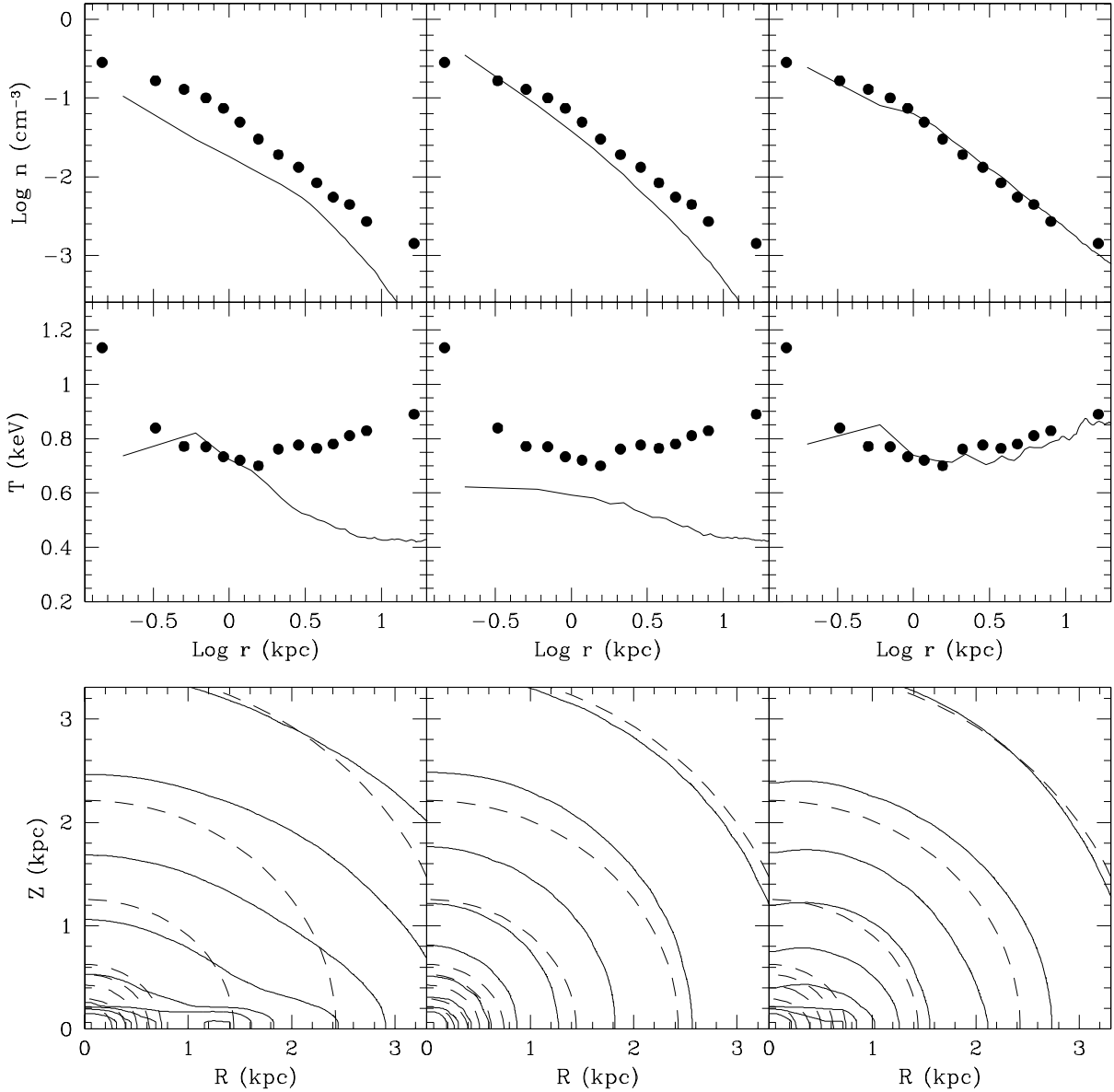}
\caption{
Two dimensional rotating flows for NGC 4649 computed after 6 Gyrs 
in three columns of panels: 
{\it Left panels:} flow without turbulent 
viscosity or circumgalactic gas;
{\it Central panels:} flow with turbulent 
viscosity but no circumgalactic gas;
{\it Right panels:} flow without turbulent 
viscosity but with circumgalactic gas.
The upper two panels compare computed, azimuthally averaged 
gas temperature and density with observations from 
Humphrey et al. (2008).
Panels at the bottom compare the 
computed bolometric X-ray isophotes 
with the observed X-ray ellipticity $\epsilon_X(r)$. 
({\it dashed lines}).
}
\label{f4}
\end{figure}

\clearpage
\begin{figure}[ht]
\vskip-2.0in
\centering
\includegraphics[bb=250 216 422 769,scale=1.0,angle=0]{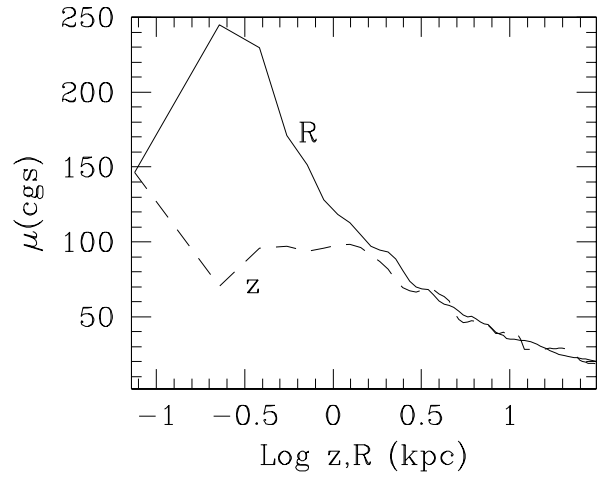}
\caption{
Turbulent viscosity profiles after 6 Gyrs 
along the $z$-axis ($\mu(0,z)$; {\it dashed line})
and along the $R$-axis ($\mu(R,0)$; {\it solid line})
both using reference model parameters
($v_t = 50$ km s$^{-1}$ and $f = 0.05$).
The viscosity is evaluated at the center of each 
computational zone and therefore must agree for 
the innermost zone.
}
\label{f5}
\end{figure}

\clearpage
\begin{figure}[ht]
\vskip-2.0in
\centering
\includegraphics[bb=250 216 422 769,scale=1.0,angle=0]{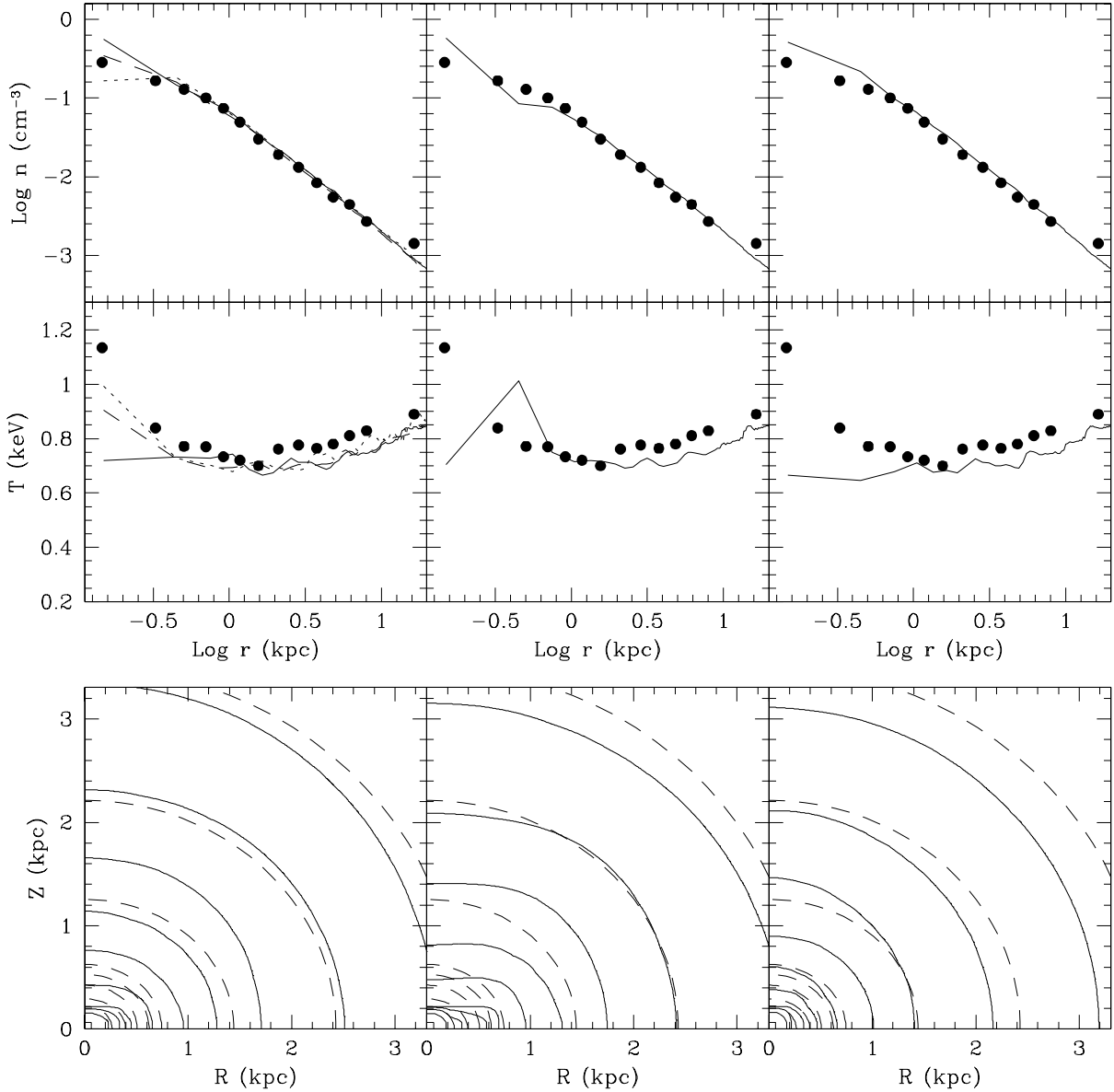}
\caption{
Two dimensional flows for NGC 4649 computed after 6 Gyrs
in three columns of panels:
{\it Left panels:} reference flow with turbulent
viscosity parameters $v_t = 50$ km s$^{-1}$ and $f = 0.05$. 
The flow oscillates unrealistically near the center as gas 
is removed rather abruptly due to cooling,   
even at relatively high spatial resolution. The three curves 
show typical excursions at 
4 Gyrs ({\it dotted lines}), 
6 Gyrs ({\it solid lines}), and 8 Gyrs ({\it dashed lines});
{\it Central panels:} flow with turbulent
viscosity parameters $v_t = 25$ km s$^{-1}$ and $f = 0.025$;
{\it Right panels:} flow with turbulent
viscosity parameters $v_t = 50$ km s$^{-1}$ and $f = 0.1$.
All flows contain circumgalactic gas.
The upper two panels compare computed, azimuthally averaged
gas temperature and density with observations from
Humphrey et al. (2008).
Panels at the bottom compare the
computed bolometric X-ray isophotes
with the observed X-ray ellipticity. 
({\it dashed lines}).
}
\label{f6}
\end{figure}

\clearpage
\begin{figure}[ht]
\vskip-2.0in
\centering
\includegraphics[bb=250 216 422 769,scale=1.0,angle=0]{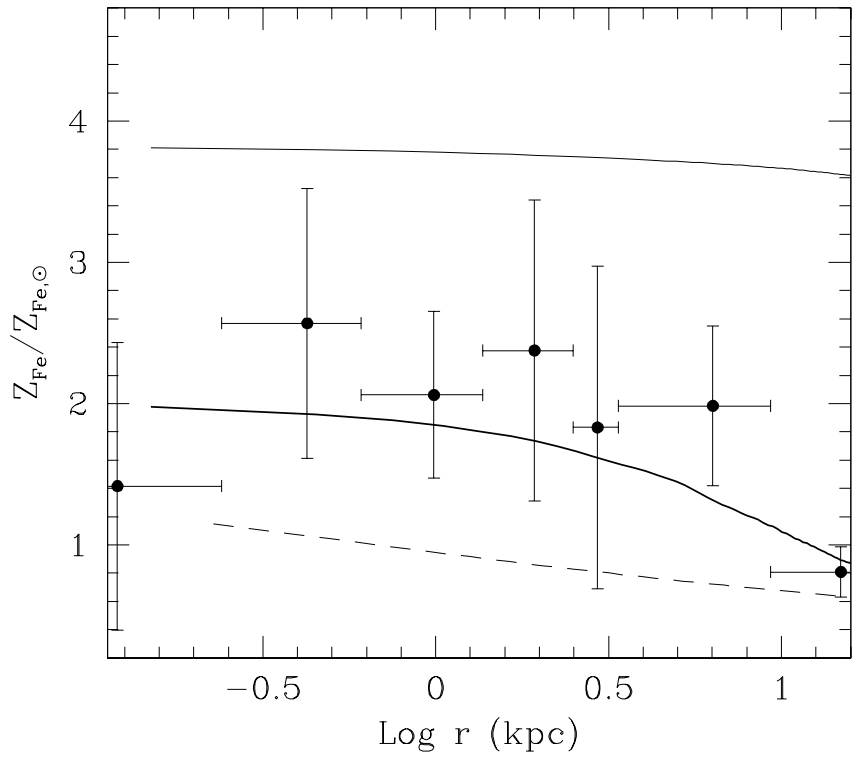}
\caption{
Plot of azimuthally averaged 
gas phase ({\it heavy solid line}) and stellar 
({\it dashed line}) iron abundance profiles after 6 Gyrs
for the reference model with circumgalactic gas having 
initial abundance  $z_{cgg} = 0.6z_{\odot}$.
The light solid line shows the unrealistically high 
abundances that result in the reference flow 
when circumgalactic gas is absent. 
The hot gas iron abundance observed in NGC 4649 shown with error bars 
are based on data discussed in Humphrey et al. (2007).
}
\label{f7}
\end{figure}

\clearpage
\begin{figure}[ht]
\vskip-2.0in
\centering
\includegraphics[bb=250 216 422 769,scale=1.0,angle=0]{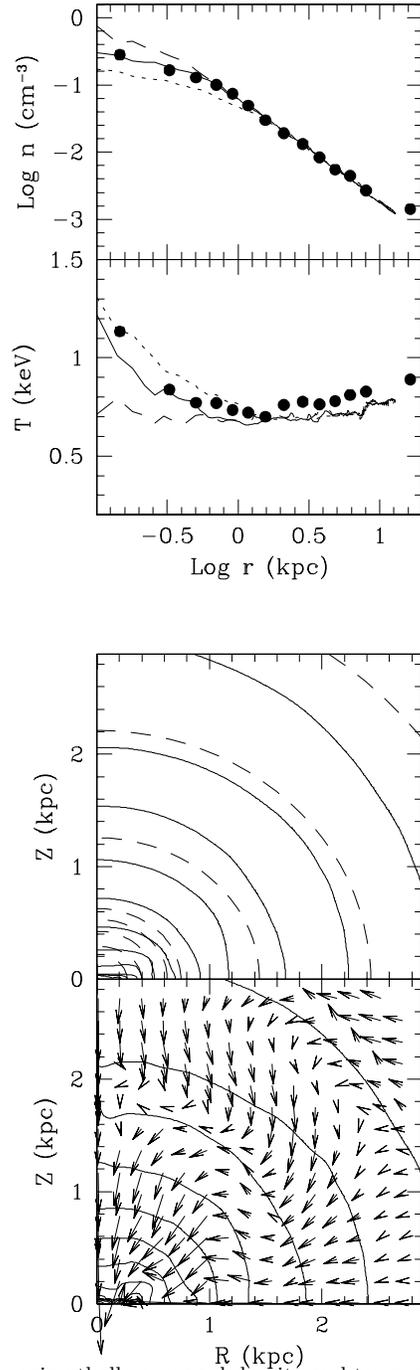}
\vskip.5in
\caption{
The top two panels show 
azimuthally-averaged density and temperature profiles for the 
the reference flow 
($v_t = 50$ km s$^{-1}$ and $f = 0.05$)
computed at high resolution ({\it solid lines}).
The two additional high resolution flows 
are done with different viscosity parameters:
$v_t = 25$ km s$^{-1}$ and $f = 0.025$, ({\it dotted lines}), and
$v_t = 50$ km s$^{-1}$ and $f = 0.1$, ({\it dashed lines}).
The third panel down compares the 
computed bolometric X-ray contours ({\it solid lines}) 
with observations ({\it dashed lines}).
The bottom panel shows the velocity field in the 
hot gas. 
The velocity in km s$^{-1}$ can be found by 
multiplying the vector length in kpc by 75.
}
\label{f8}
\end{figure}

\clearpage
\begin{figure}[ht]
\vskip-2.0in
\centering
\includegraphics[bb=250 216 422 769,scale=1.0,angle=0]{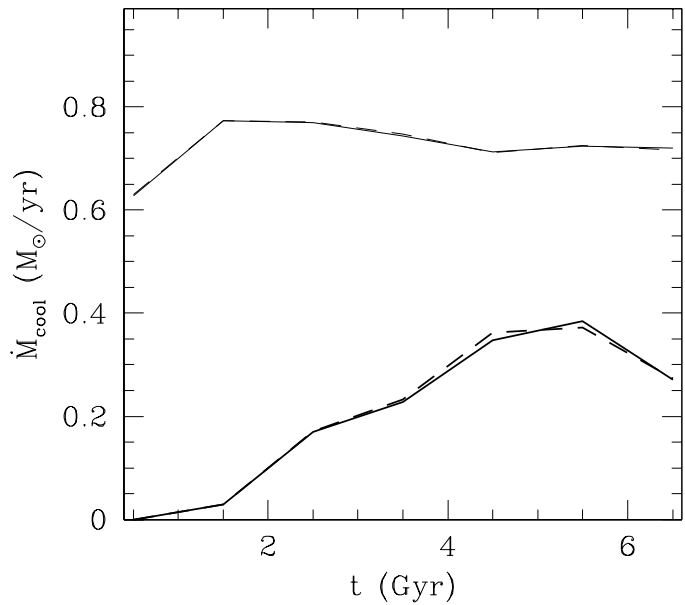}
\caption{
Rate that gas would cool near the center of NGC 4649.
{\it Heavy lines}: Flows without circumgalactic gas without 
viscosity ({\it dashed heavy line}) and with viscosity 
parameters $v_t = 50$ km s$^{-1}$ and $f = 0.1$ 
({\it solid heavy line}).
{\it Light lines}: Flows with circumgalactic gas without
viscosity ({\it dashed light line}) and with viscosity
parameters for the reference model 
$v_t = 50$ km s$^{-1}$ and $f = 0.05$
({\it solid light line}). 
These two cooling rates are nearly identical.
}
\label{f9}
\end{figure}

\end{document}